\documentclass[aps,pra,epsfig]{revtex4}

\usepackage[dvips]{graphics}
\usepackage{graphicx}
\usepackage{amsfonts}
\usepackage{amssymb}
\usepackage{amsmath}

\begin{document}

\title{Rabi switch of condensate wavefunctions in a 
multicomponent Bose gas}
\author{H.E. Nistazakis$^1$, Z. Rapti$^2$,
D.J. Frantzeskakis$^1$,  P.G. Kevrekidis$^3$, P. Sodano$^4$
\footnote{Permanent address: 
Dipartimento di Fisica and Sezione INFN, Universit\`a di Perugia, Via A. Pascoli, I-06123, Perugia, Italy}, and
A. Trombettoni$^5$} 
\affiliation{
$^1$ Department of Physics, University of Athens, Panepistimiopolis, Zografos, Athens 15784, Greece \\
$^2$ Department of Mathematics, University of Illinois at Urbana-Champaign, Urbana, Illinois 61801-2975 \\
$^3$ Department of Mathematics and Statistics, University of Massachusetts, Amherst MA 01003-4515, USA \\
$^4$ Max-Planck Institut f\"ur Physik Komplexer Systeme, N\"othnitzer Str. 38, 01167, Dresden, Germany \\
$^5$ International School for Advanced Studies and Sezione INFN, Via Beirut 2/4, I-34104, Trieste, Italy }

\begin{abstract}
Using a time-dependent linear (Rabi) coupling between the components of a weakly interacting 
multicomponent Bose-Einstein condensate (BEC), we propose a protocol for transferring the wavefunction of one component
to the other. This ``Rabi switch'' can be generated in a binary BEC mixture by an electromagnetic field 
between the two components, typically two hyperfine states. When the wavefunction to be transfered 
is - at a given time - a stationary state of the 
multicomponent Hamiltonian, then, after a time delay (depending on the 
Rabi frequency), it is possible to have the {\em same} wavefunction on the other condensate. 
The Rabi switch can be used to transfer also moving bright matter-wave solitons, as well as vortices and vortex lattices
in two-dimensional condensates. 
The efficiency of the proposed switch is shown to be 
100 $\%$ when inter-species and intra-species interaction strengths are equal. 
The deviations from equal interaction strengths are analyzed 
within a two-mode model and the dependence of the efficiency on the interaction strengths  
and on the presence of external potentials is examined in both $1D$ and $2D$ settings. 
\end{abstract}
\maketitle

\section{Introduction}

The past decade has witnessed a tremendous explosion of interest 
in the experimental and theoretical studies of Bose-Einstein condensates (BECs) 
\cite{books,reviews}. Numerous aspects of this novel and experimentally 
accessible form of matter have been since then intensely studied; one 
of them concerns the investigation of the behavior of multicomponent BECs, which have been 
experimentally studied in either mixtures of different spin states of 
$^{23}$Na \cite{stamper,leanhardt03,dumke06} 
or $^{87}$Rb \cite{myatt,dsh,williams00,minardi00,smerzi03,mandel03,chang04,wheeler04,schweikhard04,higbie05,sadler06,lundblad06,
kronjager06,vengalattore07,mertes07},
or even in mixtures of different atomic species such as 
$^{41}$K-$^{87}$Rb \cite{KRb,minardi07} and $^{7}$Li-$^{133}$Cs \cite{LiCs}.

The dynamics of a 
multicomponent BEC is described, at the mean-field level, by coupled Gross-Pitaevskii (GP) equations, 
taking into account the self- and cross- interactions between the species. In this 
framework, a number of properties and interesting phenomena have already been extensively 
analyzed. Among them, one can list ground state solutions 
\cite{shenoy,pu,esry} and small-amplitude excitations \cite{excit} of the order
parameters in multicomponent BECs, as well as 
the formation of domain walls \cite{Marek} 
and various types of matter-wave soliton complexes 
\cite{epjd}, spatially periodic states \cite{decon} and modulated 
amplitude waves \cite{mason}. Quantum phase transitions in Bose-Bose mixtures 
have 
been investigated both theoretically \cite{chen03,kuklov04,zheng05,pai08} 
and experimentally \cite{minardi07}. 
Moreover, several relevant works analyzed different aspects of 
purely spinor ($F=1$) condensates 
(which have been created in the experiments \cite{stamper,chang04}), including 
the formation of spin textures \cite{leanhardt03}, 
spin domains \cite{spindomain}, 
various types of vector matter-wave solitons \cite{wadati06,malomed,hector,beata07}, studies of 
ferromagnetic properties 
\cite{saito}, and so on.

An important resource for the experimental control of 
multicomponent BECs is the possibility to use a two-photon transition
to transfer an arbitrary fraction of atoms from one component to another, e.g., from the $\vert 1,-1 \rangle $ 
spin state of $^{87}$Rb to the $\vert 2,1 \rangle$ state. The transfer can also occur by using an electromagnetic field 
inducing a linear coupling, proportional to the Rabi frequency, 
between the different components. In Ref. \cite{decon} it was shown that, in 
analogy with 
systems arising in the field of nonlinear fiber optics (such as a twisted fiber with two linear 
polarizations, or an elliptically deformed fiber with circular polarizations \cite{opt}), 
exact Rabi oscillations between two condensates can be analytically found when inter-species coupling are equal to unity
(in proper dimensionless units). 

In this paper, we propose a protocol enabling the transfer of 
the wavefunction of a condensate to another, even in presence of interactions. The proposed protocol requires a time-dependent 
Rabi frequency: this ``Rabi switch'' is realized by turning-on the linear coupling for a pertinent period of time, 
so as to transfer the maximal fraction of the condensate from the first to the second component. The efficiency of the switch is maximal, 
if all interaction strengths (nonlinearity coefficients) are equal. If one deviates from the ideal case, 
the efficiency is modified: to analyze more realistic situations, we show that it is possible to effectively 
describe the deviations from equal interaction strengths by a two-mode ansatz, where the impossibility of transferring 
all the particles from a condensate to the other is 
identified as the self-trapping of the initial condensate wavefunction. 
Even though in the original experiments (see e.g. \cite{myatt}) the Rabi coupling was used
to transfer ground states between two repulsive condensates, our protocol 
can be efficiently used 
for transferring also 
matter-wave solitons in one-dimensional (1D) attractive condensates, as well as vortices and even vortex lattices
in two-dimensional (2D) repulsive condensates. 
We study the efficiency
of the proposed Rabi switch in each of these situations and we discuss the generalization of the same idea to a $3$-component 
condensate, where our approach would realize a ``Rabi router'' of matter into desired components.

The protocol proposed in this paper would allow for to copy a wavefunction 
from a condensate to the other in the presence of either attractive or
repulsive 
interactions among atoms, and this could improve the 
efficiency in the experimental manipulation of matter solitons and vortices; from this point of view, it provides a matter-wave 
counterpart for optical switches realized in nonlinear fiber optics, which are important tools to control optical solitons \cite{agrawal}.

Our presentation is structured as follows. In Section II we present
the theoretical framework needed to describe the Rabi switch in a two-component Bose gas, 
which is valid for attractive or repulsive interactions.
In Section III the generalization to multicomponent BECs is discussed. 
In Sections IV and V we provide the results of our analysis in 
1D and 2D settings, respectively; 
there, we also show how the external trapping potentials affect the efficiency of the Rabi switch, and 
compare the findings of the two-mode model with numerical results. 
Finally, in Section VI we present the conclusions and outlook of this work. 

\section{The Rabi Switch}

The prototypical system we consider is a two-component Bose gas in an external trapping potential: 
typically the condensates are different Zeeman levels of alkali atoms 
like $^{87}$Rb. 
Experiments with a two-component $^{87}$Rb condensate use atom states customarily denoted by 
$\vert 1 \rangle$ and $\vert 2 \rangle$; in particular,  
the states can be $\vert F=2,m_F=1 \rangle$ and $\vert 2,2 \rangle$,  
like, e.g., in \cite{smerzi03}, or $\vert 1,-1 \rangle$ and 
$\vert 2,1 \rangle$, like, e.g., in \cite{williams00} (see also the recent work \cite{mertes07}). 
In general, the condensates $\vert 1 \rangle$ and $\vert 2 \rangle$ have 
different magnetic moments: then in a magnetic trap they can be 
subjected to different magnetic potentials $V_1$ and $V_2$, eventually 
centered at different positions and having the same frequencies (like in the setup described 
in \cite{williams00}) or different frequencies \cite{smerzi03}. In \cite{smerzi03}, the ratio of the frequencies of 
$V_2$ and $V_1$ is $\sqrt{2}$. It is also possible to add a 
periodic potential acting on the two-component 
Bose gas \cite{mandel03,minardi07}. 

\begin{figure}[t]
\begin{center}
\includegraphics[width=6.cm,height=8.cm,angle=270,clip]{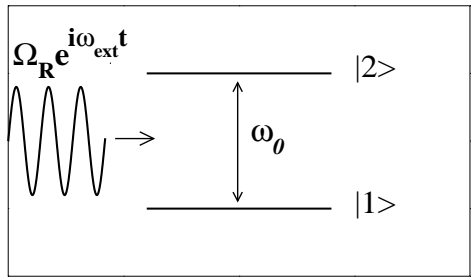} 
\caption{Josephson coupling of the two Zeeman levels $\vert 1 \rangle$ and 
$\vert 2 \rangle$ through an electromagnetic field with frequency $\omega_{\rm ext}$ 
and strength characterized by the Rabi frequency $\Omega_R$.} 
\label{intro}
\end{center}
\end{figure}

The two Zeeman states $\vert 1 \rangle$ and $\vert 2 \rangle$ 
can be coupled by an electromagnetic field 
with frequency $\omega_{\rm ext}$ and strength characterized by the 
Rabi frequency $\Omega_R$, as schematically shown in Fig. \ref{intro}. A discussion of (and references on) the experimental 
manipulation of multicomponent Bose gases can be found 
in \cite{ketterle,emergent}
(for a recent experimental realization of this coupling, 
see also \cite{mertes07}). 
The detuning is defined as 
$\omega_{\rm ext}-\omega_0$, where $\hbar \omega_0$ is the 
energy splitting between the two states 
(e.g., in \cite{smerzi03} $\omega_0 \sim 2\pi \times 2$ MHz). For concreteness we assume that the Rabi 
coupling can be turned on starting at a given instant, say $t_0 \geq 0$; at later times, the Rabi coupling 
coherently transfers particles between $\vert 1 \rangle$ and $\vert 2 \rangle$ at a Rabi frequency 
$\Omega_R$. When the number of components is larger than two, more coupling electromagnetic fields could similarly be added. 
The transfer of particles between hyperfine levels may be also 
regarded as an ``internal Josephson effect'', 
since it is similar to the Josephson tunneling of particles between Bose condensates 
in a double-well potential \cite{smerzi97,albiez05}; the only difference is that in the ``internal Josephson effect''
the two condensates are spatially overlapping, while the left and right 
part of a single-species BEC 
in a double-well are separated by the energy barrier. 
Thus, the roles of the Rabi frequency and the detuning in the 
internal Josephson effect are analogous to the ones played by the tunneling rate and the difference between 
zero-point energies of the two wells, respectively. 

In the rotating wave approximation, the dynamics of the two-component Bose-Einstein condensates is described  by 
two coupled GP equations \cite{cirac98,villain99,williams00}, 
which, in a general $3D$ setup and in dimensionless units, read 
\begin{eqnarray}
i \frac{\partial \psi_{1}}{\partial t}=
\left[-\frac{1}{2} \Delta+V_1(\vec{r})+ g_{11}\vert \psi_{1}\vert ^{2}+
g_{12}\vert \psi _{2}\vert ^{2}\right]\psi_{1}+ \alpha(t) \psi_{2},
\label{lceqa} \\
i \frac{\partial \psi_{2}}{\partial t}=
\left[-\frac{1}{2} \Delta+V_2(\vec{r})+ g_{12}\vert \psi _{1}\vert ^{2}+
g_{22}\vert \psi _{2}\vert ^{2}\right]\psi_{2}+ \alpha(t) \psi_{1},
\label{lceqb}
\end{eqnarray}
where $\psi_j(\vec{r},t)$ are the wavefunctions of the $j$-th condensate ($j=1,2$), 
$V_j$ are the 
respective trapping potentials (typically, an 
harmonic potential and/or an 
optical lattice, plus the eventual detuning, absorbed in them) and 
the quantities $g_{ij}$, which are proportional to 
the scattering lengths $a_{ij}$ of the interactions
between the species $i$ and the species 
$j$, describe the intra- ($j=i$) and 
inter- ($j \neq i$) species interactions. The system (\ref{lceqa})-(\ref{lceqb}) consists of two
linearly and nonlinearly coupled GP equations: the linear coupling is provided by the Rabi field, while 
the nonlinear coupling is proportional to $g_{12}$ and is due to the scattering
between particles of the different species. 
The scattering lengths $a_{ij}$ for experiments with Zeeman levels of $^{87}$Rb atoms are 
normally quite similar: in fact, if $\vert  1 \rangle = \vert 2,1 \rangle$ and 
$\vert 2 \rangle=\vert 1,-1 \rangle$ the scattering length ratios are $a_{11}:a_{12}:a_{22} = 0.97:1.00:1.03$, while if 
$\vert 1 \rangle=\vert 2,1 \rangle$ and $\vert 2 \rangle=\vert 2,2 \rangle$ they are
$a_{11}:a_{12}:a_{22} = 1.00:1.00:0.97$. Furthermore, one of the $a_{ij}$'s can be varied 
through Feshbach resonances \cite{books}. The term $\alpha$ is proportional 
to the Rabi frequency $\Omega_R$ \cite{nota1}, and serves the purpose of transferring the condensate wavefunction of 
$\vert 1 \rangle$ to condensate $\vert 2 \rangle$ and of controlling the time modulation of the Rabi frequency.


When the external potentials are the same ($V_1=V_2 = V$) and the 
interactions strengths are equal ($g_{11}=g_{22}=g_{12}=g$), 
Eqs. (\ref{lceqa})-(\ref{lceqb}) can be rewritten in a more compact form as:
\begin{equation}\label{eqn:stark}
i \frac{\partial {\psi}}{\partial t}  =
-\frac{1}{2}\Delta {\psi}+({\psi}^\dagger G \psi)
\psi+V({\bf r})\psi + \alpha(t) P \psi,
\end{equation}
where
\begin{equation}
\label{psi:G}
\psi=\left( 
\begin{array}{c}
\psi_1\\
\psi_2
\end{array}
\right), ~~
G=g\left(
\begin{array}{cc}
1&0\\
0&1
\end{array}
\right), ~~ 
P=\left(
\begin{array}{cc}
0&1\\
1&0
\end{array}
\right).
\end{equation}
As a result of the fact that $G$ and $P$ commute, one can decompose 
the solution $\psi$ of the ``inhomogeneous'' problem described by Eq. (\ref{eqn:stark}) as
\begin{eqnarray}
\psi(\vec{r},t)=U(t) \phi(\vec{r},t),
\label{stark1}
\end{eqnarray}
where $U(t)$ is the matrix of the homogeneous problem
\begin{equation}\label{unit}
U(t)=\exp{\left[ -i P {\cal I}(t)  \right]} = \left(
\begin{array}{cc}
\cos{{\cal I}(t)}& -i \sin{{\cal I}(t)}\\
-i \sin{{\cal I}(t)} & \cos{{\cal I}(t)}
\end{array}
\right),
\end{equation}
with ${{\cal I}(t)}=\int_0^t\alpha(t') dt'$. Substituting Eq. (\ref{unit}) in Eq. (\ref{eqn:stark}), 
it is readily found that $\phi(\vec{r},t)$ satisfies the evolution equation \cite{decon}
\begin{equation}
\label{eqn:nls}
i \frac{\partial \phi}{\partial t} \equiv {\cal H} \phi=-\frac{1}{2} \Delta \phi +(\phi^\dagger G \phi) \phi+
V(\vec{r}) \phi,
\end{equation}
which is identical to Eq. (\ref{eqn:stark}), but without the Rabi term proportional to $\alpha(t) P$. 

When the external potentials or the interaction strengths are different, it is formally possible, 
as discussed in the Appendix A, to perform a decomposition like the one given in Eq. (\ref{stark1}) and remove the Rabi term. 
However, this is done at the price of introducing
time- and space- dependent effective interaction strengths in the nonlinear terms: in particular, 
for different external potentials the effective interaction strengths are both time- and space- dependent, 
while when the interaction strengths are different, 
a nonlinear Josephson term also arises in the nonlinear terms (i.e., $i \dot{\phi}_1$ is proportional to 
$\phi_2$ through terms proportional to products $\phi_i^\ast \phi_j$). 
As a result, the removal of the Rabi term is of little practical use and one has to resort to numerical 
or variational estimates: in the following, we will focus on the situation of different $g_{ij}$'s, and 
show that the efficiency of the Rabi switch for small deviations from the equal strengths situation can be effectively 
described by a two-mode model. 

When the Rabi frequency is fixed, oscillations of atoms between the two components
have been studied \cite{williams99,villain99,ohberg99,williams00,smerzi03,decon,merhasin05} 
and experimentally observed \cite{williams00,smerzi03}. In the following,  
we consider a time-dependent Rabi frequency $\alpha(t)$: through a proper choice of a $\alpha(t)$, 
we can transfer the wavefunction of $\vert 1 \rangle$ to $\vert 2 \rangle$. More precisely, 
we propose a way to perform the following operation: at a time $t_0$, one has all the particles 
in $\vert 1 \rangle$ in the wavefunction $\psi_1(\vec{r},t_0)$, and no particles 
in $\vert 2 \rangle$ ($\psi_2(\vec{r},t_0)=0$). 
At a time $t_1$ we wish to have all the particles in $\vert 2\rangle$ in the same 
wavefunction: $\psi_2(\vec{r},t_1)=\psi_1(\vec{r},t_0)$ (apart a phase factor). The protocol proposed in this paper 
allows the transfer of the wavefunction if $\psi_1(\vec{r},t_0)$ is a stationary state (or a moving soliton, 
as discussed in Section IV) of the nonlinear Hamiltonian ${\cal H}$ defined in (\ref{eqn:nls}). 
If it is not, we can however have 
at the time $t_1$ all the particles in $\vert 2 \rangle$ in the wavefunction 
the condensate $\vert 1 \rangle$ would have had without the Rabi coupling [see Eq. (\ref{gen})]. 

The proposed protocol works with or without nonlinearity: 
but with the nonlinearity on, one can transfer also a soliton wavefunction; for instance, 
in $1D$ one can have a matter-wave soliton of the species $\vert 1 \rangle$ 
propagating with velocity $v$, and after a time delay, the proposed Rabi switch will generate the same 
soliton with the same velocity in the species $\vert 2 \rangle$. 
Two remarks are due: (i) the transfer mechanism has the highest
possible efficiency for equal interaction strengths, 
however it is very good and its efficiency is close to $1$ 
in a wider region in the relevant parameter space [in $1D$, deviations from the integrable case of equal interaction 
strengths are discussed through a two-mode ansatz]; (ii) the proposed mechanism is not copying the full many-body 
wavefunction of the weakly interacting Bose gas, but only the order parameter, which is related to the one-body density matrix.

To be more specific, we assume that $\alpha(t)$ depends on time as 
\begin{equation}
\label{timedep}
\alpha(t)= \left \{\begin{array}{ll}
0,       &  \hspace{10mm}       0 \leq t  < t_0, \\
\gamma,  &  \hspace{10mm}       t_0 \leq t \leq t_1 = t_0 + \delta, \\
0,       &  \hspace{10mm}       t > t_1,
        \end{array}
\right.
\end{equation}
where $t_0$ is the switch-on time and $\delta$ denotes the duration of the Rabi pulse. 
From Eq. (\ref{timedep}) we readily find 
%
\begin{equation}
\label{timedep:I}
{\cal I}(t)= \left \{\begin{array}{ll}
0,       &  \hspace{10mm}       0 \leq t  \leq t_0, \\
\gamma (t-t_0),  &  \hspace{10mm}       t_0 \leq t \leq t_1 , \\
\gamma \delta,       &  \hspace{10mm}       t \geq t_1.
\end{array}
\right.
\end{equation}

Introducing the 
vector field $\phi$ by the decomposition in Eq. (\ref{unit}), 
it is observed that at the time $t_0$, i.e., {\em before} the switch-on of the Rabi pulse, 
$\psi(\vec{r},t_0)=\phi(\vec{r},t_0)$, while 
for $t_0 \leq t \leq t_1$ we find that 
%
\begin{equation}
\left \{ \begin{array}{ll}
\psi_1(\vec{r},t) = \cos{[\gamma (t-t_0)]} \phi_1(\vec{r},t)-i \sin{[\gamma (t-t_0)]} \phi_2(\vec{r},t), \\
\psi_2(\vec{r},t) = -i \sin{[\gamma (t-t_0)]} \phi_1(\vec{r},t) + \cos{[\gamma (t-t_0)]} \phi_2(\vec{r},t).
        \end{array}
\right.
\label{lceqb:t}
\end{equation}
When the pulse duration is
\begin{equation}
\label{pulse}
\delta= \frac{\pi}{2 \gamma},
\end{equation}
at the end of the pulse we obtain 
%
\begin{equation}
\left \{ \begin{array}{ll}
\psi_1(\vec{r},t_1) = -i \phi_2(\vec{r},t_1), \\
\psi_2(\vec{r},t_1)= -i \phi_1(\vec{r},t_1).
\end{array}
\right.
\label{endb}
\end{equation}
Since, in the interval $[t_0,t_1]$, $\phi$ satisfies the homogeneous coupled GP 
equations (\ref{eqn:nls}), i.e. the same equations satisfied by the vector field $\psi$ in the interval 
$[0,t_0]$, then 
\begin{equation}
\left( 
\begin{array}{c}
\psi_2(\vec{r},t_1)\\
\psi_1(\vec{r},t_1)
\end{array}
\right)=
-ie^{-i {\cal H} (t_1-t_0)} 
\left( 
\begin{array}{c}
\psi_1(\vec{r},t_0)\\
\psi_2(\vec{r},t_0)
\end{array}
\right).
\label{gen}
\end{equation}
Notice that if, instead of Eq. (\ref{timedep}), one allows for a different time dependence of $\alpha(t)$, the time $t_1$ at which 
Eq. (\ref{gen}) holds is given by the condition $\cos{{\cal I}(t_1)}=0$. For instance, 
with $\alpha(t)=0$ for $t < t_0$ and $t > t_0+\delta$, 
and $\alpha(t)=f(t)$ for $t_0 \leq t \leq t_0+\delta$, the pulse duration $\delta$ such that Eq. (\ref{gen}) is valid 
is given by $\int_0^\delta dt' f(t'-t_0) =\pi/2$.

We are interested in the situation in which no particles are in $\vert 2 \rangle$ at $t_0$ 
($\psi_2(\vec{r},t_0)$=0); this situation may occur, e.g., in the case where only a single condensate has been prepared
(if, eventually, particles 
exist in the other component, it is possible to remove them by the suitable application of a Rabi pulse).
Notice that this has been experimentally realized e.g. in the experiments
of \cite{mertes07} (see also references therein). 
In such a case, Eq. (\ref{gen}) 
implies that at the end of the pulse one has that (apart from a phase factor) the wavefunction 
describing the condensate $\vert 2 \rangle$ is 
the same wavefunction which the condensate $\vert 1 \rangle$ would have 
had in $t_1$ 
in the absence of the Rabi pulse. We can quantify the success of the described protocol in several ways: 
one of them will be to define the ``efficiency'' $T$ as the fraction of atoms we are able to transfer from 
$\vert 1 \rangle$ to $\vert 2 \rangle$, i.e., 
\begin{equation}
T= \frac{N_2(t_1)}{N_1(t_0)}, 
\label{efficiency}
\end{equation}
where $N_i(t)=\int d\vec{r} \vert \psi_i \left( \vec{r},t \right) \vert^2$ is the number 
of particles in the condensate $i$ at time $t$. 
Notice that such a definition can even be extended in cases where the
number of atoms in the second component is not zero initially
by replacing in the numerator of Eq. (\ref{efficiency}) $N_2(t_1) \rightarrow
(N_2(t_1)-N_2(t_0))$.
Another more stringent way is to define a kind of 
``fidelity'' $F$ of the wavefunction transfer, i.e., the quantity
\begin{equation}
F=\int d\vec{r} \, \vert \psi_2^\ast(\vec{r},t_1) \vert \cdot \vert \psi_1(\vec{r},t_0) \vert.
\label{fidelity}
\end{equation}
From Eq. (\ref{gen}) we see that the efficiency of Rabi switch is $1$ for equal interaction strengths, but the fidelity is not. 
However, we can have the fidelity to be equal to $1$ 
if $\psi$ is a stationary state of ${\cal H}$ corresponding to the eigenvalue 
$\mu$, i.e. 
\begin{equation}
{\cal H} \psi_\mu(\vec{r}) = \mu \psi_\mu(\vec{r}).
\label{stat}
\end{equation}
If at time $t=0$, $\psi(\vec{r},0)=\psi_\mu(\vec{r})$, then $\phi(\vec{r},t_0)=e^{-i \mu t_0} \psi_\mu(\vec{r})$ and 
\begin{equation}
\left \{\begin{array}{ll}
\psi_1(\vec{r},t_1)= e^{-i \mu \delta} \left[ \cos{(\gamma \delta)} \, \psi_1(\vec{r},t_0) - 
i \sin{(\gamma \delta)} \, \psi_2(\vec{r},t_0)
\right], \\
\psi_2(\vec{r},t_1)= e^{-i \mu \delta} \left[ - i \sin{(\gamma \delta)} \, \psi_1(\vec{r},t_0) + 
\cos{(\gamma \delta)} \, \psi_2(\vec{r},t_0) \right].
\end{array}
\right.
\label{end:stat}
\end{equation}
When no particles are in $\vert 2 \rangle$ at $t_0$ and the pulse duration is given by Eq. (\ref{pulse}), one has
\begin{equation}
\left \{\begin{array}{ll}
\psi_1(\vec{r},t_1)= 0, \\
\psi_2(\vec{r},t_1)= -i e^{-i \mu \delta} \psi_1(\vec{r},t_0).
\end{array}
\right.
\label{endb1:stat}
\end{equation}
Equations (\ref{endb1:stat}) show that the wavefunctions of the two components have been exchanged up to a phase factor. 
This remarkable feature allows us, again with equal interaction strengths $g_{ij}$, to transfer the condensate wavefunction 
(with $100 \%$ efficiency) from a populated hyperfine state to an empty one. In the following section we discuss how to 
transfer from a condensate to the other the wavefunction of a moving matter-wave soliton.

We should further note about the latter that the nature of linear operators in the right hand-side of
Eq. (\ref{eqn:stark}) is irrelevant in the derivation of Eq. (\ref{stark1}). 
Hence, our analysis can be used to deal with:
\begin{itemize}
\item repulsively interacting as well as attractively interacting
systems;
\item continuum, as well as discrete systems;
\item homogeneous systems (in the absence of external potentials)
or inhomogeneous systems 
(e.g., in the presence of external 
harmonic trap and/or optical lattice potentials);
\item one-dimensional systems or higher-dimensional ones.
\end{itemize}

In what follows, we illustrate the versatility of the Rabi switch 
by examining characteristic examples for each of the above settings. 
We will then illustrate, how 
the perfect efficiency of the matter wave transfer 
(discussed above for equal inter-particle interactions) is ``degraded''
in more realistic situations (where such interactions are no longer equal).

\section{Generalization to Multicomponent Bose-Einstein Condensates}

The protocol discussed in the previous Section can be generalized for 
${\cal N} \geq 2$ components with a suitable choice of the time dependence 
of the Rabi frequencies 
$\alpha_{ij}$ transferring particles from the condensate $i$ to the condensate $j$. 
For instance, for ${\cal N}=3$ and for equal potentials ($V_1=V_2=V_3 \equiv V$) and interaction strengths 
($g_{ij} \equiv g$), the relevant system of the three coupled GP equations can be written in the form of Eq. (\ref{eqn:stark}), namely 
\begin{equation}\label{eqn:stark:3}
i \frac{\partial {\psi}}{\partial t}  =
-\frac{1}{2}\Delta {\psi}+({\psi}^\dagger G \psi)
\psi+V({\bf r})\psi + \tilde{P}(t) \psi
\end{equation}
with
\begin{equation}
\psi=\left( 
\begin{array}{c}
\psi_1\\
\psi_2 \\
\psi_3
\end{array}
\right), ~~
G=g \left(
\begin{array}{ccc}
1&0&0\\
0&1&0\\
0&0&1
\end{array}
\right), ~~
\tilde{P}=\left(
\begin{array}{ccc}
0&\alpha_{12}(t) &\alpha_{13}(t)\\
\alpha_{12}(t)&0&\alpha_{23}(t)\\
\alpha_{13}(t)&\alpha_{23}(t)&0
\end{array}
\right).
\end{equation}
For general $\alpha_{ij}$, the decomposition $\psi=U \phi$ with 
\begin{equation}
U=e^{- i \int_0^t \tilde{P}(t') dt'}
\label{U-3}
\end{equation}
fails to recast Eq. (\ref{eqn:stark:3}) in the homogeneous form 
\cite{footnote};
however, it still removes the Rabi term when $\alpha_{ij}(t)= \alpha(t)$ for any $i,j$. With 
$\psi=U \phi$ and $U$ given by Eq. (\ref{U-3}), 
one finds $i \frac{\partial {\phi}}{\partial t}  =
-\frac{1}{2}\Delta {\phi}+({\phi}^\dagger G \phi) \phi$. The matrix $U_{ij}(t)$ ($i,j=1,2,3$) has  
diagonal elements $U_{jj}=(1/3)\left[ 2 \exp\left(i {\cal I} \right) + \exp \left(-2 i {\cal I} \right) \right]$ 
and off-diagonal ones $U_{ij}=U_{jj}- \exp\left(i {\cal I}\right)$. Once  
the Rabi term has been removed, 
the ``Rabi switch'' described in the previous Section can be applied also for general ${\cal N}$ to transfer a wavefunction 
from a condensate to any one of the others. 

Another choice of $\alpha_{ij}$ allowing for the removal of the Rabi term 
is provided by the generalization of 
Eq. (\ref{timedep}), namely
\begin{equation}
\label{timedep:N}
\alpha_{ij}(t)= \left \{\begin{array}{ll}
0,       &  \hspace{10mm}       0 \leq t  \le t_0, \\
\gamma_{ij},  &  \hspace{10mm}       t_0 \leq t \leq t_1, \\
0,       &  \hspace{10mm}       t \ge t_1,
        \end{array}
\right.
\end{equation}
with all the $\alpha_{ij}$ turned on/off at the same time, but with eventually different intensities. 
As an example, for ${\cal N}=3$, one may consider $\gamma_{12}=a_1$, $\gamma_{13}=a_2$ and $\gamma_{23}=0$: 
the matrix $U(t)$, for $t_0 \leq t \leq t_1$, then reads
\begin{equation}
\label{3comp1}
U(t)=
\left(
\begin{array}{ccc}
r_1 (1+r_2)&
a_1 r_1 \frac{1-r_2}{\sqrt{a_1^2+a_2^2}}&
a_2 r_1 \frac{1-r_2}{\sqrt{a_1^2+a_2^2}} \\
a_1 r_1 \frac{1-r_2}{\sqrt{a_1^2+a_2^2}}&
\frac{a_2^2+a_1^2 r_1 (1+r_2)}{a_1^2+a_2^2}&
a_1 a_2 r_1 \frac{(r_3-1)^2}{a_1^2+a_2^2}\\
a_2 r_1 \frac{1-r_2}{\sqrt{a_1^2+a_2^2}} &
a_1 a_2 r_1 \frac{(r_3-1)^2}{a_1^2+a_2^2}&
\frac{a_1^2+a_2^2 r_1 (1+r_2)}{a_1^2+a_2^2}
\end{array}
\right),
\end{equation}
where 
\begin{eqnarray}
2 r_{1}=r_{2}^{-1/2}=r_{3}^{-1}=\exp\left(-i \sqrt{a_1^2+a_2^2} \left(t-t_0 \right) \right).
\end{eqnarray}
This allows us to determine the transfer of matter from the first
to the second and third component.
Similar results may be obtained in the more general case of ${\cal N} \ge 2$ components, i.e., 
``desirable'' amounts of matter can be controllably directed to different hyperfine states 
according to their Rabi couplings. This general ``Rabi router'' 
is quite interesting in its own right, and it could be experimentally implemented in $F=1$ spinor condensates   
\cite{stamper,chang04,higbie05}.

\section{Results for $1$D settings}

We now consider the $1$D version of Eqs. (\ref{lceqa})-(\ref{lceqb}), which 
is relevant to the analysis of 
``cigar-shaped'' condensates confined in highly anisotropic traps \cite{books,reviews}, and in dimensionless units reads
\begin{eqnarray}
i \frac{\partial \psi _{1}}{\partial t}=
\left[-\frac{1}{2} \frac{\partial^2}{\partial x^2}+V_1(x)+ g_{11}\vert \psi_{1}\vert ^{2}+
g_{12}\vert \psi _{2}\vert ^{2}\right]\psi_{1}+ \alpha(t) \psi_{2},
\label{lceqa:1D} \\
i \frac{\partial \psi_{2}}{\partial t}=
\left[-\frac{1}{2} \frac{\partial^2}{\partial x^2} +V_2(x)+ g_{12}\vert \psi _{1}\vert ^{2}+
g_{22}\vert \psi _{2}\vert ^{2}\right]\psi_{2}+ \alpha(t) \psi_{1}.
\label{lceqb:1D}
\end{eqnarray}
We use wavefunctions $\psi_i(x,t)$ normalized to 
unity, so that $N_1(t)+N_2(t)=1$, with $N_i(t)=\int dx \vert \psi_i(x,t) 
\vert^2$. When the effect of the external potentials $V_i$ is negligible (as, e.g., in the case of potentials varying slowly 
on the soliton scale) and in absence of the Rabi coupling 
($\alpha=0$), the system (\ref{lceqa:1D})-(\ref{lceqb:1D}) becomes the 
Manakov system \cite{manakov}, which is integrable 
for $g_{11}=g_{12}=g_{22}$. In what follows we examine both attractive and 
repulsive interatomic interactions, corresponding, respectively, to negative and positive values of
the scattering lengths, and we will consider the effect of the presence of the 
trapping potential on the wavefunction transfer. 


\subsection{Stationary bright matter-wave solitons} 

\begin{figure}[t]
\begin{center}
\includegraphics[width=6.cm,height=8.cm,angle=270,clip]{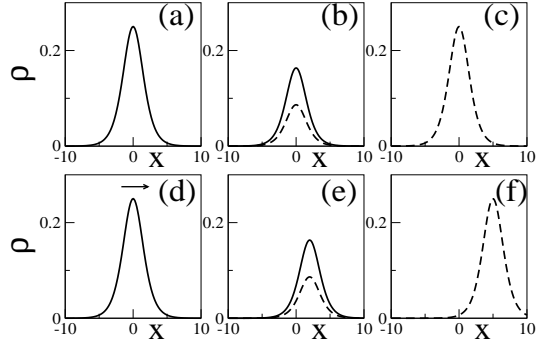} 
\caption{Transferring a stationary bright matter-wave soliton: 
in (a)-(b)-(c) the density $\rho_j=\vert \psi_j \vert^2$ is plotted for both 
components
($\vert 1 \rangle$ solid line; $\vert 2 \rangle$ dashed line) at the times 
$t=t_0,t_0+0.4 \delta, t_1$. In (d)-(e)-(f) the density is plotted at the same times, 
but for a velocity $v=1$ ($\delta=\pi/2$, $t_1=5$).} 
\label{fig-Rabi}
\end{center}
\end{figure}

We consider in this subsection attractive interactions, $g_{ij} < 0$, 
in absence of external potentials. 
Putting $\ell_{ij} = - g_{ij}$, 
we first consider the ideal case where 
$\ell_{11}=\ell_{12}=\ell_{22} \equiv \ell$ 
and assume that, at $t=0$, all the particles of 
$\vert 1 \rangle$ are described by the $1$-soliton solution of the nonlinear Schr\"odinger equation; thus, 
$\psi_2(x,0)=0$ and 
\begin{equation}
\label{solit}
\psi_1(x,0)=\frac{\sqrt{\ell}/2}{\cosh{(\ell x/2)}}.
\end{equation}
In these units the soliton's chemical potential is $\mu=-\ell^2/8$; turning on the 
Rabi coupling $\gamma$ at time $t_0$, and then turning it off at $t_1=t_0+\delta$, 
one gets, for $t_0 \leq t \leq t_1$, 
\begin{equation}
\psi(x,t)=\frac{\sqrt{\ell}/2}{\cosh{(\ell x/2)}} e^{-i \mu (t-t_0)}\left( 
\begin{array}{c}
\cos{\gamma(t-t_0)}\\
-i \sin{\gamma(t-t_0)}.
\end{array}
\right)
\label{rest:t}
\end{equation}
For $\delta=\pi/(2 \gamma)$ no particles are in the condensate $\vert 1 \rangle$ at $t_1$, and the soliton 
wavefunction has been transferred in $\vert 2 \rangle$, i.e., $\psi_2(x,t_1)=-(i/2) e^{-i \mu \delta} \sqrt{\ell} / 
\cosh{(\ell x/2)}$. The transfer of the soliton wavefunction 
is illustrated in Fig. \ref{fig-Rabi}(a)-(c). 


Notice that if we choose 
as initial condition 
\begin{equation}
\left( 
\begin{array}{c}
\psi_1(x,0)\\
\psi_2(x,0)
\end{array}
\right)=
\frac{\sqrt{\ell}/2}{\cosh{(\ell x/2)}}
\left( 
\begin{array}{c}
\sqrt{N_1(0)} e^{i \varphi_1(0)}\\
\sqrt{N_2(0)} e^{i \varphi_2(0)}
\end{array}
\right), 
\label{partial}
\end{equation}
i.e., two bright solitons with particle numbers $N_1(0)$, $N_2(0)$ and phase difference 
$\Delta\varphi(0)=\varphi_2(0)-\varphi_1(0)$, then, 
at $t=t_1$, we obtain 
\begin{equation}
N_1(t_1)=\left( \cos{(\gamma \delta)} \sqrt{N_1(0)}+ \sin{(\gamma \delta)} \sin{\Delta\varphi(0)}
\sqrt{N_2(0)}\right)^2+N_2(0) \sin^2{(\gamma \delta)} \cos^2{\Delta\varphi(0)}.
\end{equation}
This shows that, by choosing properly the pulse duration and the initial phase difference, one can transfer a ``desired'' part 
of the soliton wavefunction from one condensate to the other.

Let us discuss now the interesting situation of different interaction strengths:the aim there is to study the efficiency of the Rabi switch 
of the soliton wavefunction and qualitatively understand the effect 
of the deviation from the ideal case. To that effect, 
we introduce a variational two-mode ansatz and 
confine ourselves to the situation in which no particles are initially in $\vert 2 \rangle$.
For $t_0 \leq t \leq t_1$, we choose the variational 
vectorial wavefunction 
\begin{equation}
\psi_V=
\left( 
\begin{array}{c}
\psi_{v1}(x,t)\\
\psi_{v2}(x,t)
\end{array}
\right)=e^{-i \mu t}
\left( 
\begin{array}{c}
\sqrt{N_1(t)} e^{i \varphi_1(t)} \Phi_1(x)\\
\sqrt{N_2(t)} e^{i \varphi_2(t)} \Phi_2(x)
\end{array}
\right),
\label{variational}
\end{equation}
where
\begin{equation}
\label{Phi:i}
\Phi_i(x)=\frac{\sqrt{\ell_{ii}}/2}{\cosh{(\ell_{ii} x/2)}}.
\end{equation}
The variational parameters are the numbers of 
particles $N_i(t)$ and their phases $\varphi_i(t)$. The variational vector wavefunction (\ref{variational}) 
has been used in \cite{pare} to study the wavepacket dynamics for two linearly coupled nonlinear 
Schr\"odinger equations with $\ell_{11}=\ell_{22}$ and $\ell_{12}=0$. For general $\ell_{ij}$'s, the 
Lagrangian ${\cal L}=\frac{i}{2} \langle \psi_V^\dag \frac{\partial \psi_{V}}{\partial t}- 
\frac{\partial \psi_{V}^\dag}{\partial t} \psi_{V} \rangle-\langle \psi_V^{\dag}\tilde{\cal H} \psi_V\rangle$  
[where $\tilde{\cal H}$ is given by Eq. (\ref{H:tilde}) and
$\langle \rangle$  denotes spatial integration], 
is computed in Appendix B, where we show that the variational equations of motion 
for $N_1-N_2$ and $\varphi_2-\varphi_1$ are the equations of a (non-rigid) pendulum. The mass $M$ of the pendulum 
depends on the $\ell_{ij}$'s according Eq. (\ref{mass}) 
and for $\ell_{11}=\ell_{12}=\ell_{22}$ the pendulum mass is zero, allowing 
for the transfer of all the particles from a species to the other. 
When $\ell_{11} \neq \ell_{22}$, a detuning term in the 
pendulum equations is present [see Eq. (\ref{Delta_E})]. In the following, 
we will focus for simplicity on the more illuminating case 
$\ell_{11}=\ell_{22}$, with a general $\ell_{12}$.

Introducing the variables
\begin{equation}
\eta=N_1-N_2; \, \, \, \varphi=\varphi_2-\varphi_1,
\label{variables}
\end{equation}
one gets the equations of motion
\begin{equation}
\left \{ \begin{array}{ll}
\dot{\eta}=2 \gamma \sqrt{1-\eta^2} \sin{\varphi}, \\
\dot{\varphi}=-2 \gamma \frac{\eta}{\sqrt{1-\eta^2}}\cos{\varphi}+\ell_{11} \frac{\ell_{12}-\ell_{11}}{6} \eta,
        \end{array}
\right.
\label{var:t}
\end{equation}
with initial conditions $\eta(t_0)=1$ 
(i.e., all the particles initially in $\vert 1 \rangle$) 
and $\varphi(t_0)=0$. 
It is worth noticing that Eqs. (\ref{var:t}) 
are the same equations governing the tunneling 
of weakly-coupled 
BECs in a double-well 
potential \cite{smerzi97,raghavan99} 
(the only difference being that $\gamma$ corresponds to $-K$, where $K>0$ 
is the tunneling rate, which gives the same results 
for $\varphi \to \varphi+\pi$). 
Equations (\ref{var:t}) are formally identical 
to the equations for an electron in a 
polarizable medium, where a polaron is formed \cite{kenkre86}. 
Analytical solutions have been found for the discrete nonlinear 
Schr\"odinger equations describing the motion of the polaron 
between two sites of a dimer \cite{kenkre86,kenkre87}.

\begin{figure}[t]
\begin{center}
\includegraphics[width=6.cm,height=8.cm,angle=270,clip]{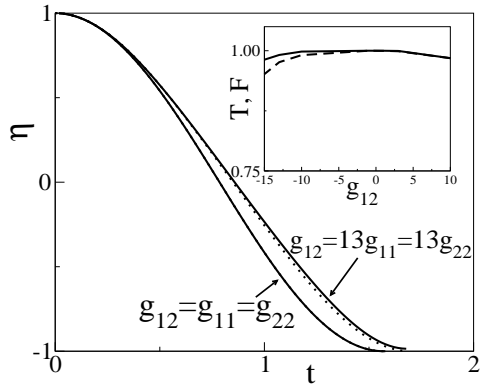} 
\caption{Comparison of the population imbalance $\eta(t)$ obtained from the numerical solution of the GP equations 
(\ref{lceqa:1D})-(\ref{lceqb:1D}) [solid line] with the results of the two-mode equations (\ref{var:t}) [dotted line] 
for $g_{12}=-1,-13$ given by Eq. (\ref{mass-pend-sempl}) to a mass pendulum $M=0,2$, respectively. Inset: 
efficiency (solid line) and fidelity (dashed line), respectively defined according to Eqs. (\ref{efficiency}) and 
(\ref{fidelity}), and obtained from the 
numerical solution of the GP equations. Parameters used in both plots: $t_0=0$, $\gamma=1$, $g_{11}=g_{22}=-1$.} 
\label{comp}
\end{center}
\end{figure}

Eqs. (\ref{var:t}) are the equations of a non-rigid pendulum 
\cite{smerzi97,raghavan99}, with the effective Hamiltonian being 
\begin{equation}
\label{Hamiltonian-pend-sempl}
H_{eff} = \frac{M}{2} \eta^2 + 2 \gamma \sqrt{1-\eta^2} \cos{\varphi}, 
\end{equation} 
where the pendulum mass is given by
\begin{equation}
\label{mass-pend-sempl}
M=\ell_{11} \frac{\ell_{12}-\ell_{11}}{6}.
\end{equation} 
When $\ell_{11}=\ell_{12}$, then the mass in Eq. (\ref{mass-pend-sempl}) vanishes and $\ddot{\eta}=-4 \gamma^2 \eta$. The duration 
$\delta $ of the pulse needed to have a perfect switch is such that $\eta(t_1=t_0+\delta)=-1$, i.e. $\delta=\pi/(2 \gamma)$ 
in agreement with Eq. (\ref{pulse}). If the mass $M$ is positive (i.e., $g_{12} < g_{11}$), then it is still possible to 
transfer all the particles from $\vert 1 \rangle$ to $\vert 2 \rangle$, provided that the mass $M$ 
is smaller than a critical value $M_c$. If $M < M_c$, then the time $t_1$ at which $\eta(t_1)=-1$ will be 
different from $t_0+\pi/(2 \gamma)$ 
(the analytical computation of the tunneling period for a mass 
$M \neq 0$ is reported in the Appendix of \cite{raghavan99}). This means that for deviations from the ideal case, one can 
optimize the transfer by choosing a pulse duration different from Eq. (\ref{pulse}). This is illustrated in Fig. \ref{comp}, 
where we compare $\eta(t)$ obtained from the two-mode equations (\ref{var:t}) with the results of the numerical solution 
of the GP equations (\ref{lceqa:1D})-(\ref{lceqb:1D}) for $\ell_{12} / \ell_{11}=13$: the numerical and variational results 
are in good agreement for a wide range of the parameters (see the inset of Fig. \ref{comp}). 
The computation of $M_c$ is done according to the method discussed in \cite{smerzi97}: 
namely, one has to compute the point at which self-trapping occurs, and the result is
\begin{equation}
\label{mass-pend-ST}
M_c=4\gamma.
\end{equation}
For $\ell_{11}=1$, the critical value of $\ell_{12}$ is equal to $25$. A comparison with the numerical solution of the GP equations 
shows that this value is overestimated: e.g., at $M=3 \gamma$, the efficiency $T$ at the optimal time is $\sim 0.9$, while it 
should be equal to $1$. However, the result (\ref{mass-pend-ST}) gives a reasonable estimate of the point at which 
is no longer possible to transfer with perfect efficiency all the particles from one condensate to the other, due to the 
self-trapping of the initial condensate wavefunction. Finally, we observe that, for $M<0$, the agreement between 
numerical and variational results is still good and the critical point is $M_c=-4 \gamma$. We also notice that similar results 
can be obtained for the Rabi switch of $N$-soliton solutions.


\subsection{Moving bright solitons and dark solitons}

The proposed protocol works also for transferring moving solitons: with 
$\ell_{11}=\ell_{12}=\ell_{22} \equiv \ell$, one can prepare the initial wavefunction as 
\begin{equation}
\psi(x,0)=\frac{\sqrt{\ell}/2}{\cosh{(\ell x/2)}} e^{i v x}\left( 
\begin{array}{c}
\sqrt{N_1(0)} e^{i \varphi_1(0)}\\
\sqrt{N_2(0)} e^{i \varphi_2(0)}
\end{array}
\right).
\label{mov:0}
\end{equation}
For $t \le t_0$ one has (with $\mu=-\ell^2/8$)
\begin{equation}
\psi(x,t)=\frac{\sqrt{\ell}/2}{\cosh{(\ell (x-vt)/2)}} e^{i (v x-\mu x-i v^2 t/2)}\left( 
\begin{array}{c}
\sqrt{N_1(0)} e^{i \varphi_1(0)}\\
\sqrt{N_2(0)} e^{i \varphi_2(0)} 
\end{array}
\right),
\label{mov:t}
\end{equation}
so that at $t_1$, i.e., at the end of the pulse, $\delta=\pi/(2 \gamma)$, one has 
%
\begin{equation}
\psi(x,t_1)=-i \frac{\sqrt{\ell}/2}{\cosh{(\ell (x-vt_1)/2)}} e^{i (v x-\mu x-i v^2 t_1/2)}\left( 
\begin{array}{c}
\sqrt{N_2(0)} e^{i \varphi_2(0)}\\
\sqrt{N_1(0)} e^{i \varphi_1(0)} 
\end{array}
\right).
\label{mov:t1}
\end{equation}
If no particles are initially in $\vert 2 \rangle$, then one can transfer the moving soliton from a condensate to the other, 
as depicted in Fig. \ref{fig-Rabi}(d)-(f). 

When the $\ell_{ij}$'s are different, one can use the same variational method discussed in the previous subsection. One needs to 
consider the variational wavefunction (\ref{variational}), but with a time-dependence included in the functions 
$\Phi$, which now read $\Phi_i(x,t)=(1/2) e^{ivx-iv^2t/2} \sqrt{\ell_{ii}}/ \cosh{[\ell_{ii}(x-vt)/2]}$: 
apart form constant terms, 
we obtain the same Lagrangian [c.f. Eq. (\ref{Lag})] and the 
analysis is the same as before. 
In particular, the threshold for the self-trapping transition is still given by Eq. (\ref{mass-pend-ST}).


On the other hand, when the parameters $g_{ij}$ are positive and equal (and $V_i=0$), 
the soliton solution is now a dark matter-wave soliton, and one can 
transfer it from one condensate to the other. To examine the situation when 
the $g_{ij}$ are different and estimate the 
self-trapping threshold, one can also use a variational approach. 
Omitting the details, when $g_{11}=g_{22}$ 
one gets $M_c=4 \gamma$, with $M=g_{11}(g_{12}-g_{11}) n$, where $n$ is the 
(asymptotic) density of the dark soliton for 
very large $x$.


\subsection{Effect of the trapping potential} 

\begin{figure}[t]
\begin{center}
\begin{tabular}{cc}
    (a) & (b) \\
\includegraphics[width=4.cm,height=5.cm,angle=0,clip]{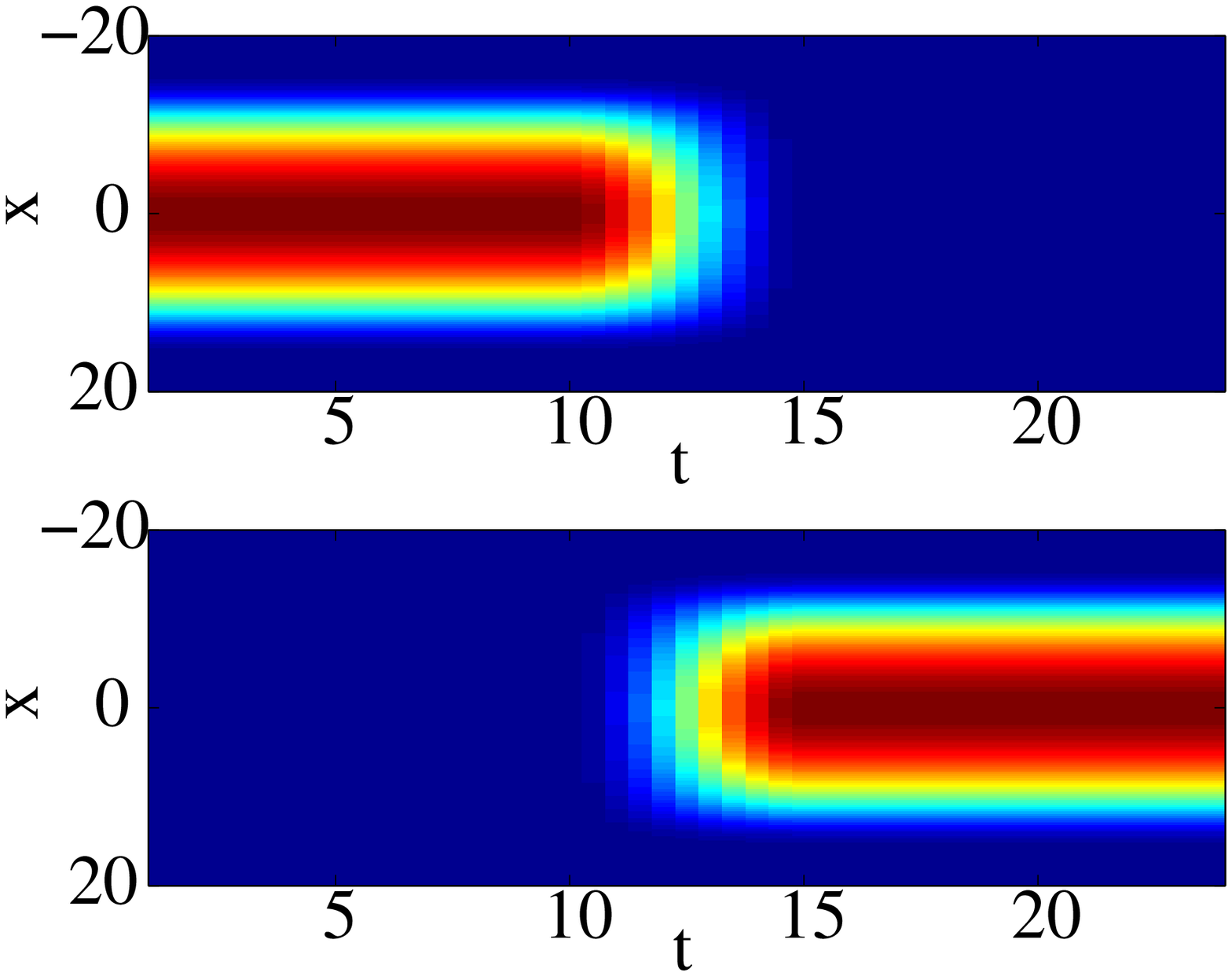} &
\includegraphics[width=4.cm,height=5.cm,angle=0,clip]{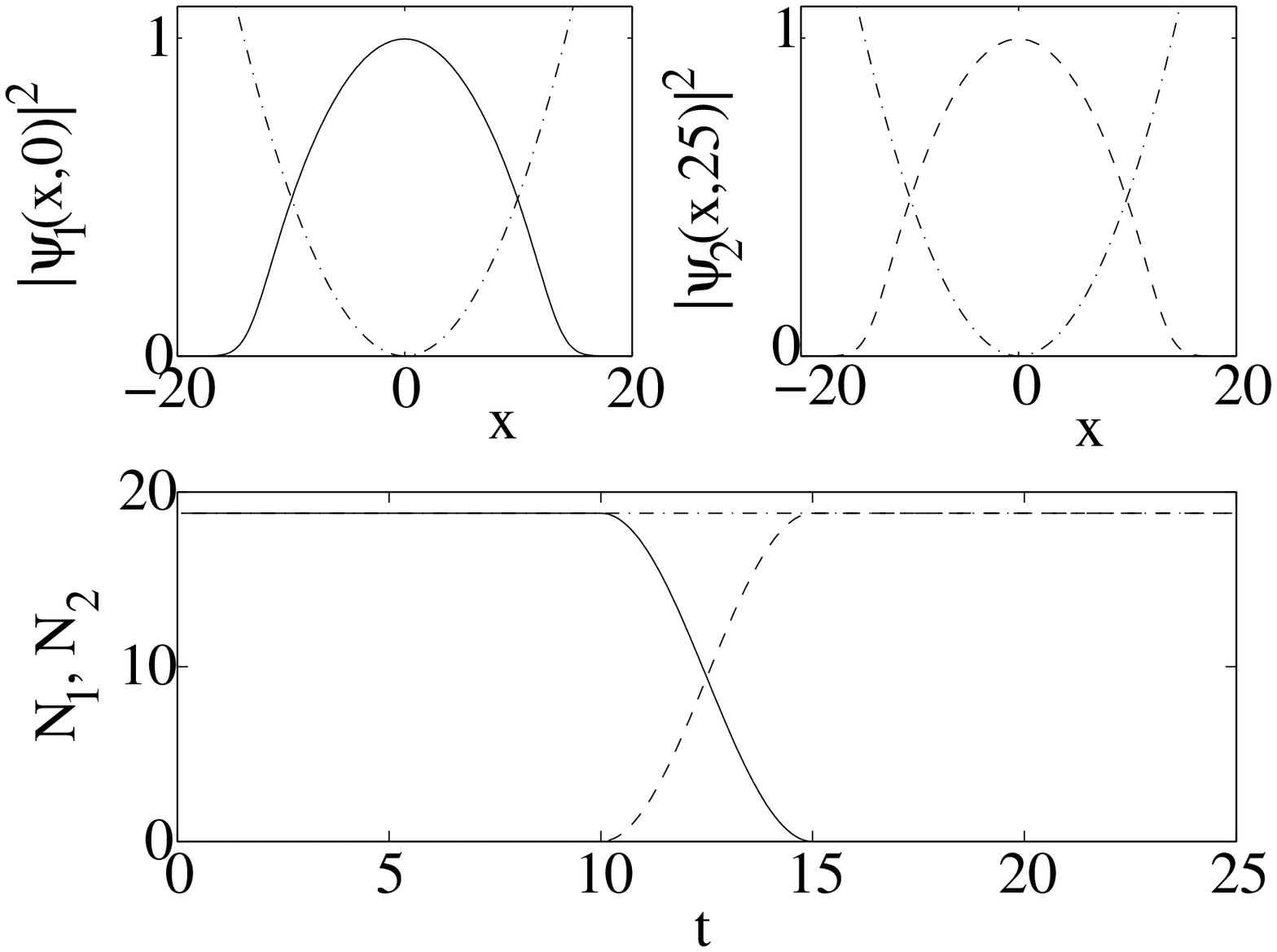} \\
    (c) & (d) \\
\includegraphics[width=4.cm,height=5.cm,angle=0,clip]{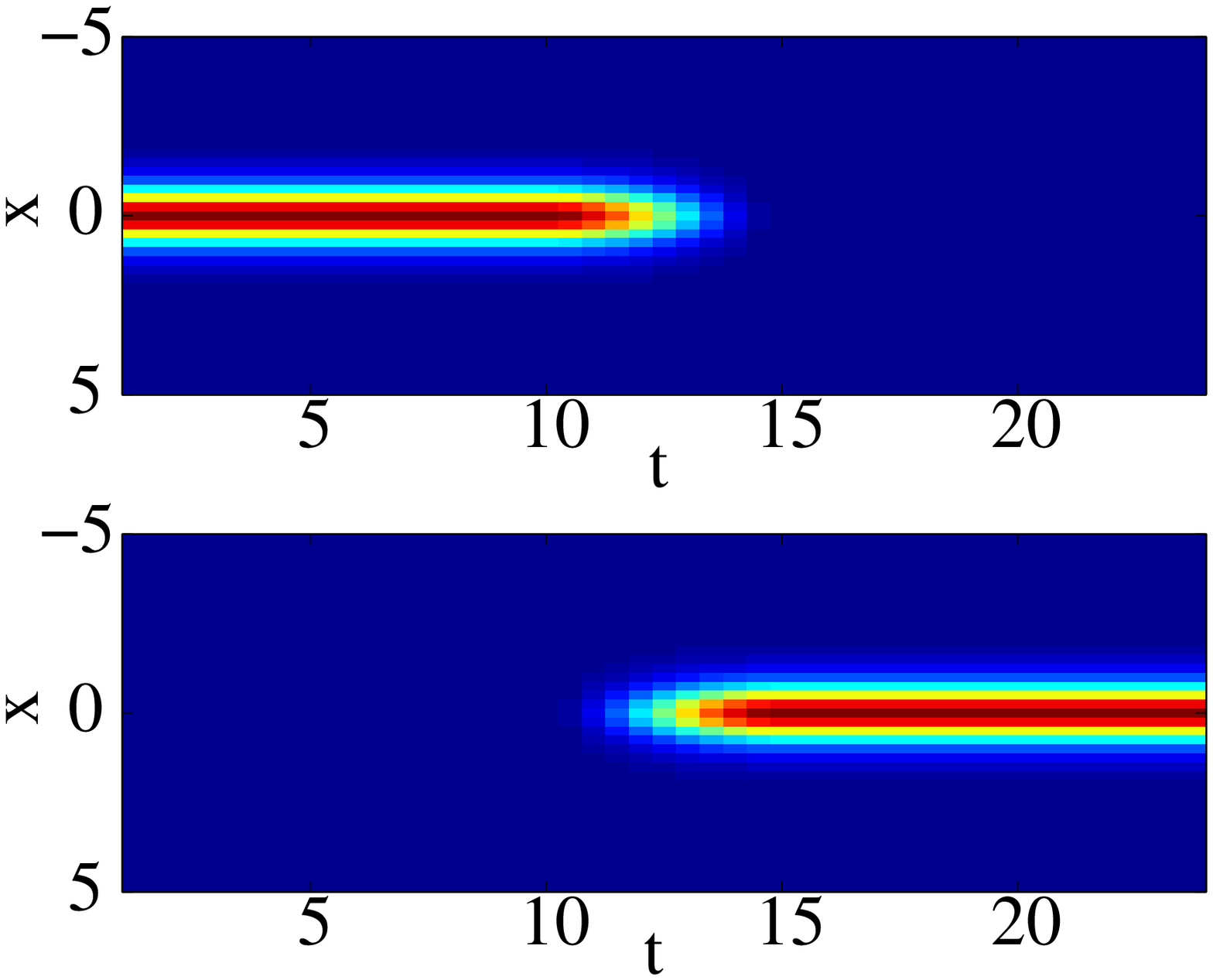} &
\includegraphics[width=4.cm,height=5.cm,angle=0,clip]{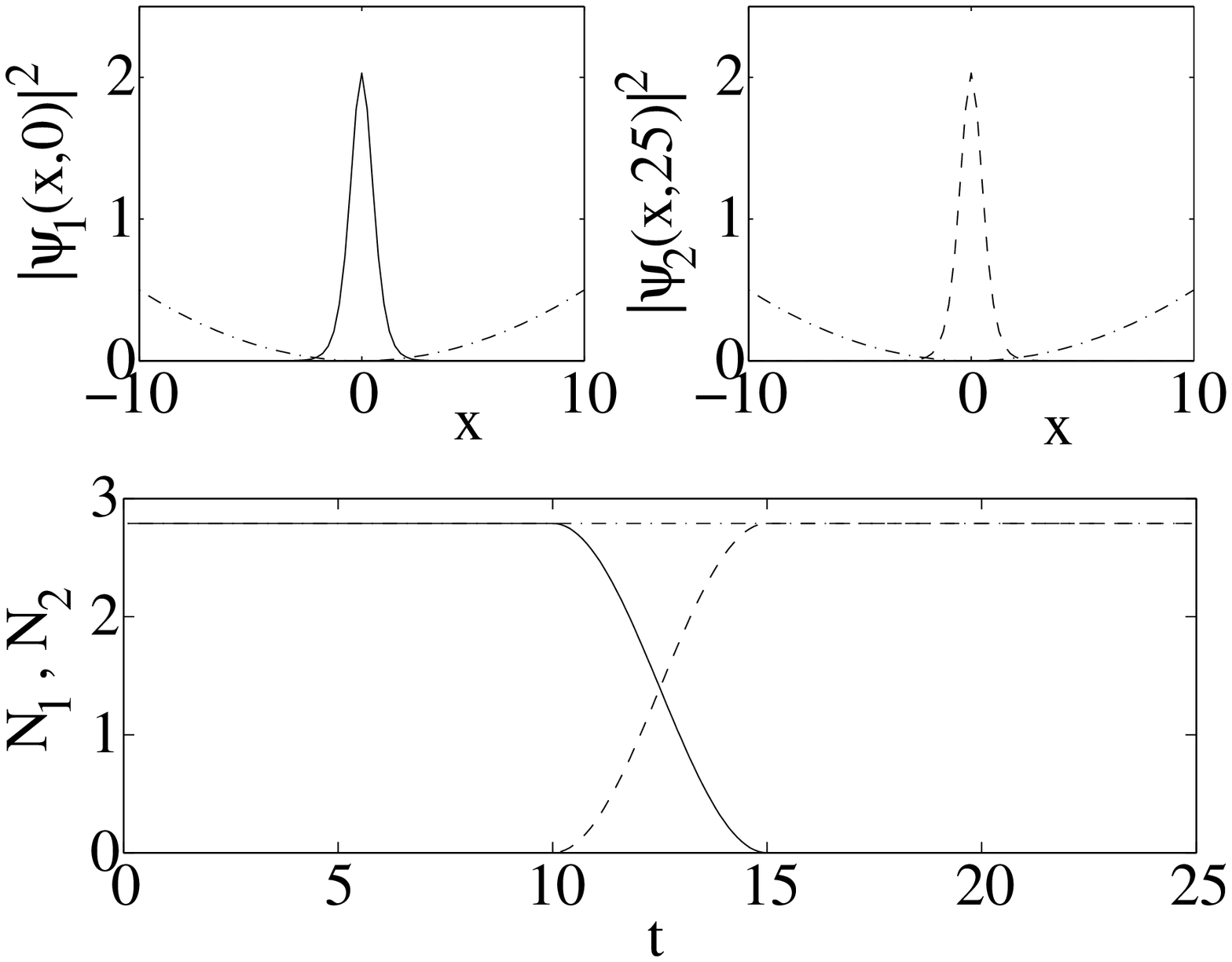} \\
 (e) & (f) \\
\includegraphics[width=4.cm,height=5.cm,angle=0,clip]{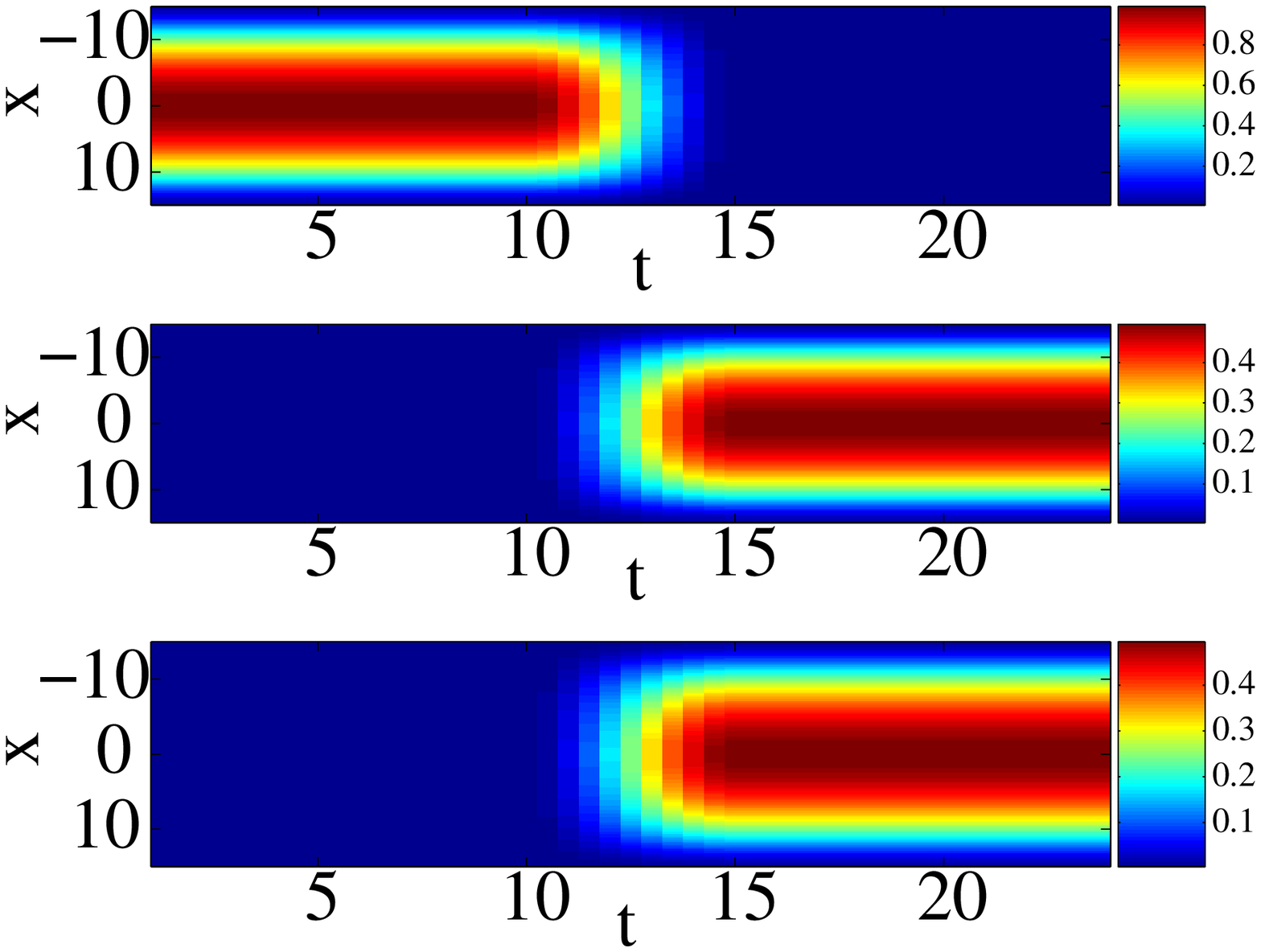} &
\includegraphics[width=4.cm,height=5.cm,angle=0,clip]{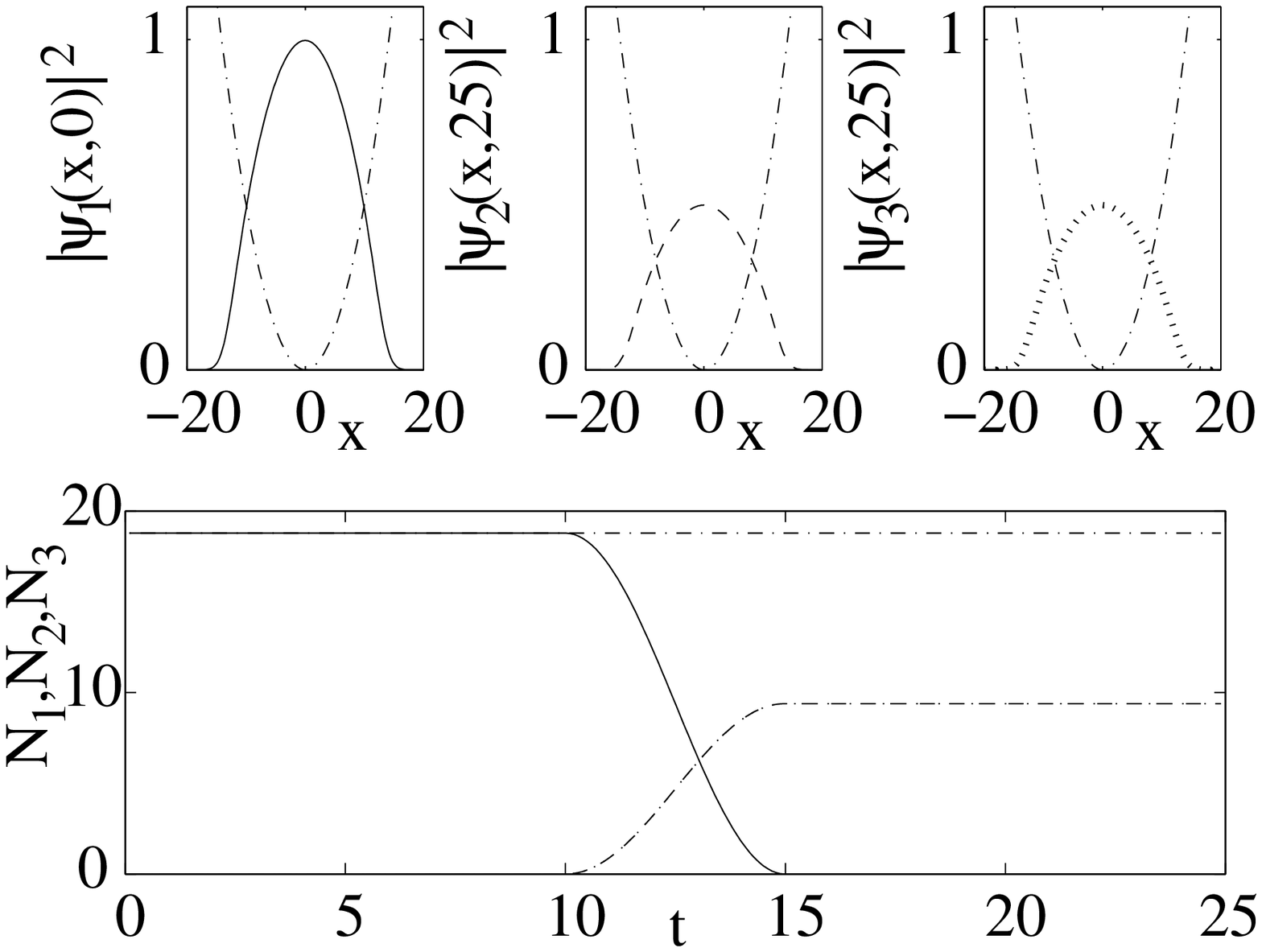} \\
\end{tabular}
\caption{Panel (a) shows a space-time plot of the density 
$\vert \psi_1(x,t)\vert ^2$ for
the first component (top panel) and $\vert \psi_2(x,t)\vert ^2$ for the
second component (bottom panel). Panel (b) shows
the spatial profile of $\vert \psi_1(x,t=0)\vert ^2$ and $\vert \psi_2(x,t=25)\vert ^2$
in solid and dashed lines respectively. The dash-dotted line shows
the magnetic trap potential. They also show the evolution of the
particle number $N_i= \int \vert \psi_i\vert ^2 dx$ for each of the components
in the interval of the dynamical evolution. These features are shown
in panels (a) and (b) for $g_{11}=g_{12}=g_{22}=1$. In 
panels (c) and (d), they are shown for a soliton in the case of 
$g_{11}=g_{12}=g_{22}=-1$. Analogous features are shown for 
$3$-component condensates in panels (e) and (f) (the third component
is shown by dotted line in the right panel), again for the case
with $g_{ij}=1$ for all $i,j=1,2,3$.}
\label{rfig1}
\end{center}
\end{figure}

First, we consider repulsive condensates, in the ideal situation where
$g_{11}=g_{12}=g_{22}=1$. An example of the realization of the Rabi switch
for the ground state of the system is shown in the top panels of Fig. \ref{rfig1}: 
we have considered an harmonic 
trapping potential of the form $V(x)=(1/2)\Omega^{2} x^{2}$. 
In the numerical simulations of the GP equations, we use $\Omega=0.08$, $t_0=10$ and $\gamma=\pi/10$; 
hence, after $t \geq 15$, the condensate wavefunctions have completely switched between components.
Next, we consider the attractive case with $g_{11}=g_{12}=g_{22}=-1$.
In this case, as an initial condition (for the first component) we have used 
a bright matter-wave soliton with the well-known sech-profile given by Eq. (\ref{solit}). 
As shown in the middle panels of Fig. \ref{rfig1}, the transfer of the wavefunction is complete, just as in the repulsive case.


To better illustrate the versatility of the Rabi switch, 
even for ${\cal N}>2$, in Fig. \ref{rfig1} 
we have also considered the case with ${\cal N}=3$. 
In accordance with the results of the analysis carried in Section III, we use $a_1=a_2=(\pi/10)/\sqrt{2}$ 
between $t_0=10$ and $t=15$. In line with Eq. (\ref{3comp1}), at the end of this time 
interval, $r_2=-1$ and, as a result (due to the symmetry in the choice of $a_1$ and $a_2$), half of the matter initially at the
first component is transferred to the second component and half to the third component, in excellent
agreement with the results shown in the bottom panel of Fig. \ref{rfig1}. 

\begin{figure}[t]
\begin{center}
\begin{tabular}{cc}
   (a) & (b) \\
\includegraphics[width=4.cm,height=5.cm,angle=270,clip]{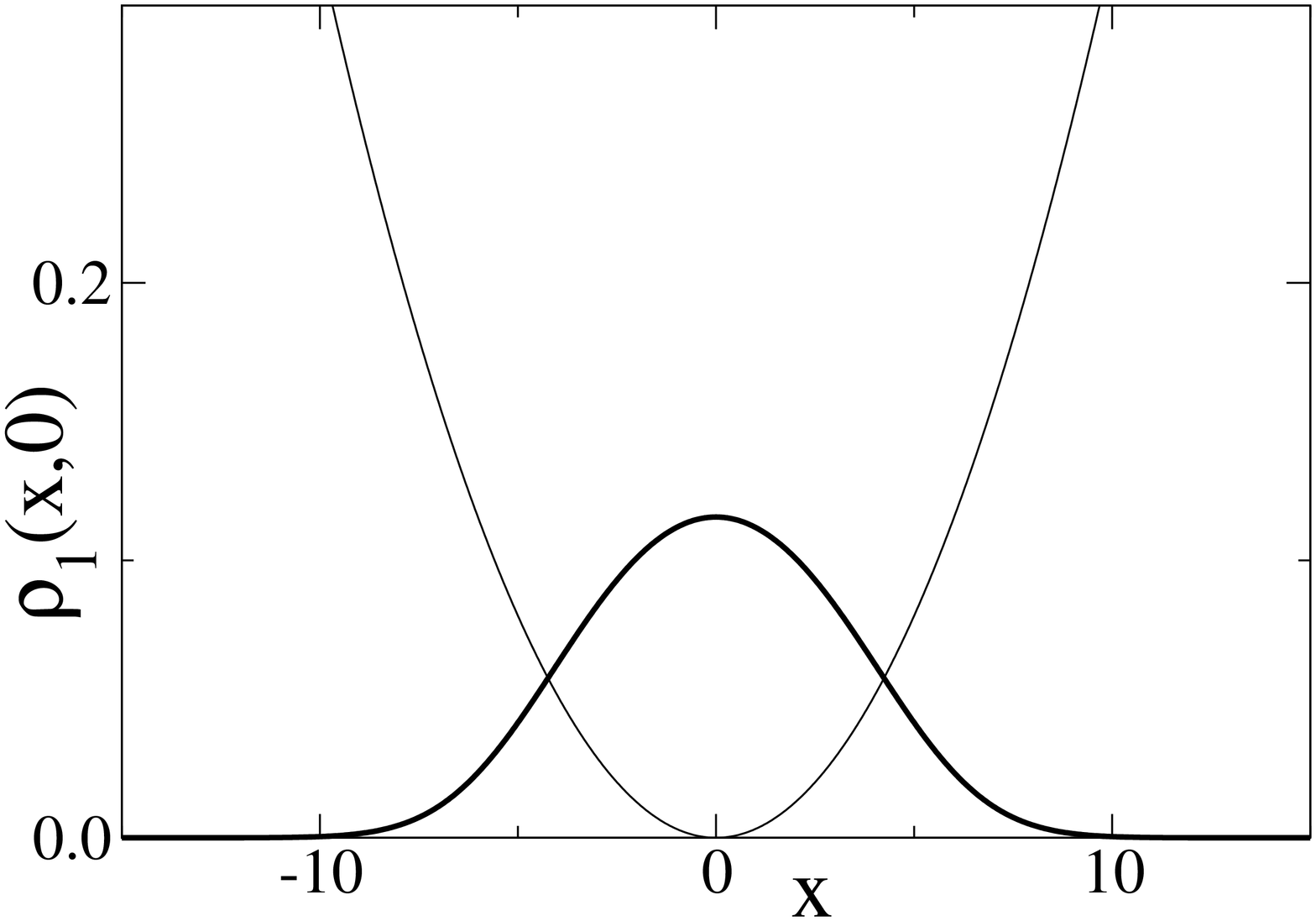} &
\includegraphics[width=4.cm,height=5.cm,angle=270,clip]{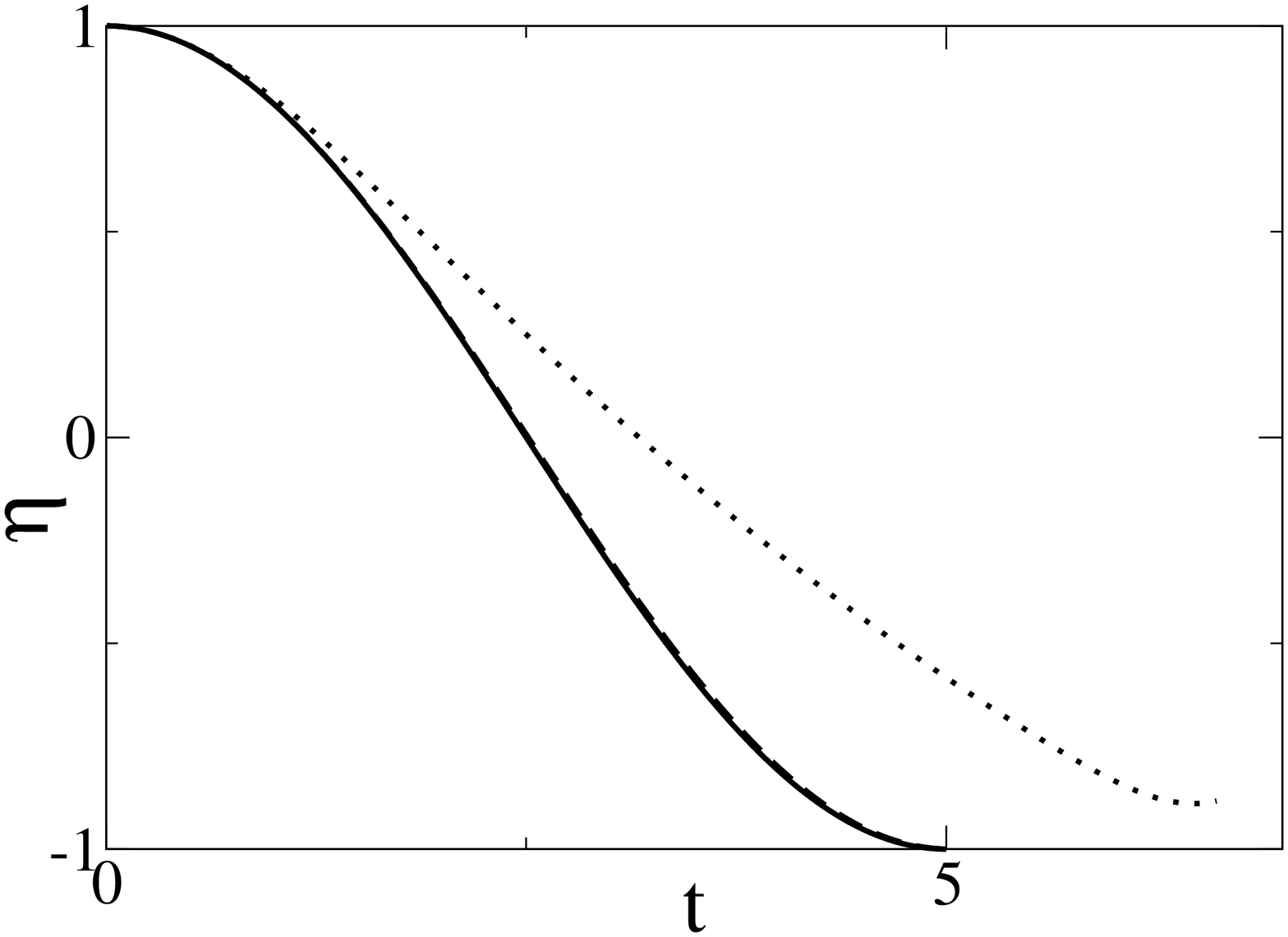} \\
   (c) & (d) \\
\includegraphics[width=4.cm,height=5.cm,angle=270,clip]{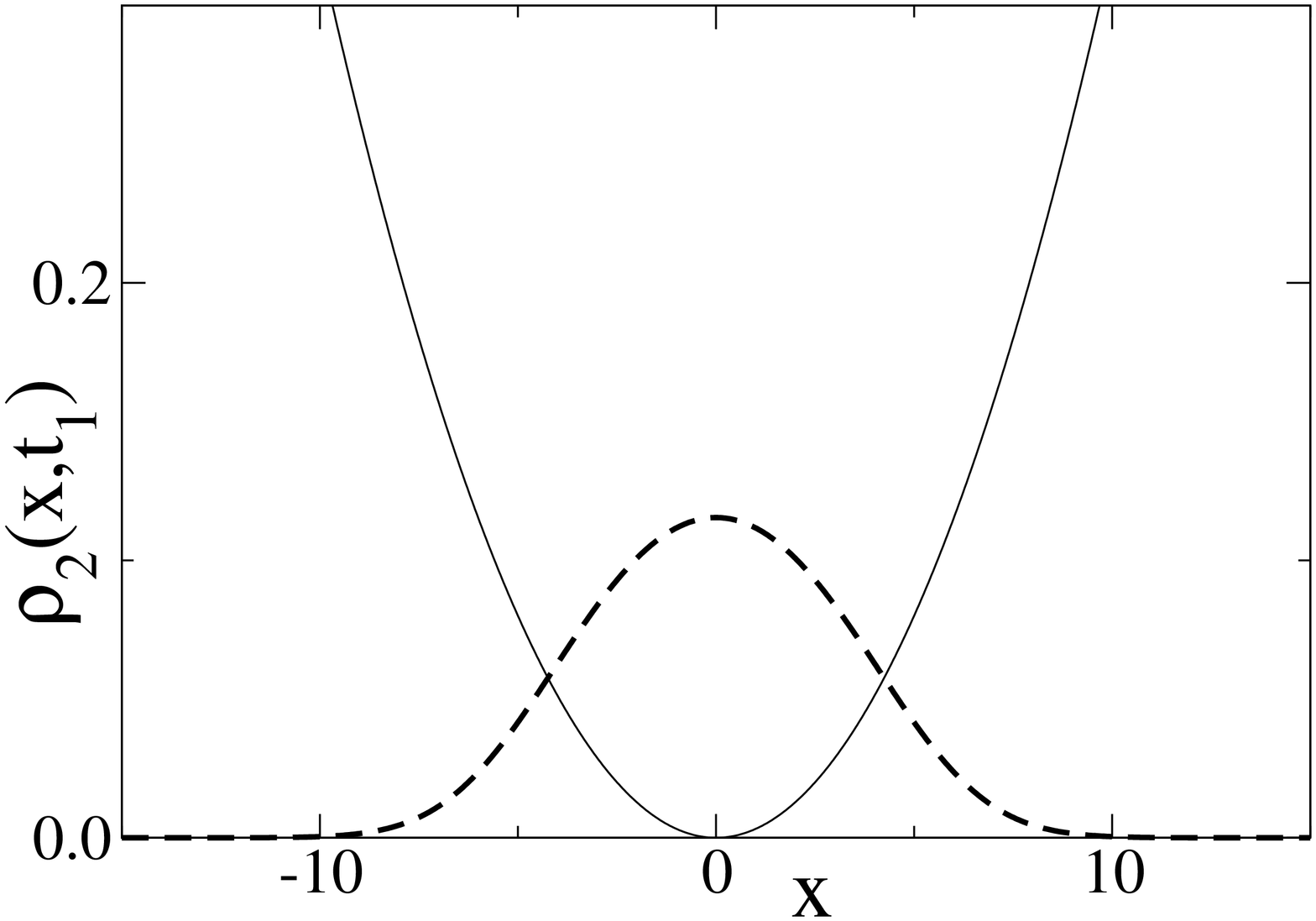} &
\includegraphics[width=4.cm,height=5.cm,angle=270,clip]{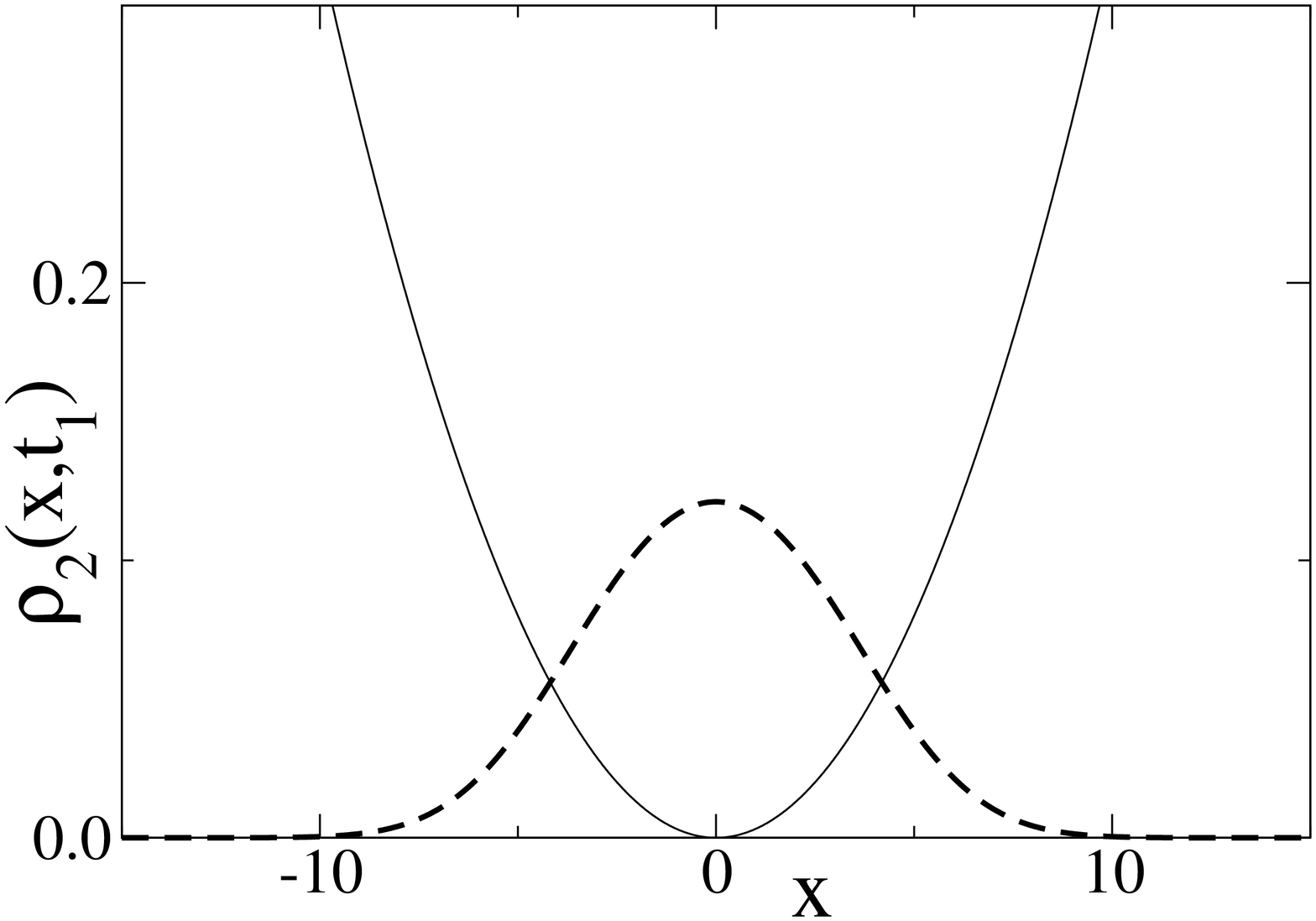} \\
   (e) & (f) \\
\includegraphics[width=4.cm,height=5.cm,angle=270,clip]{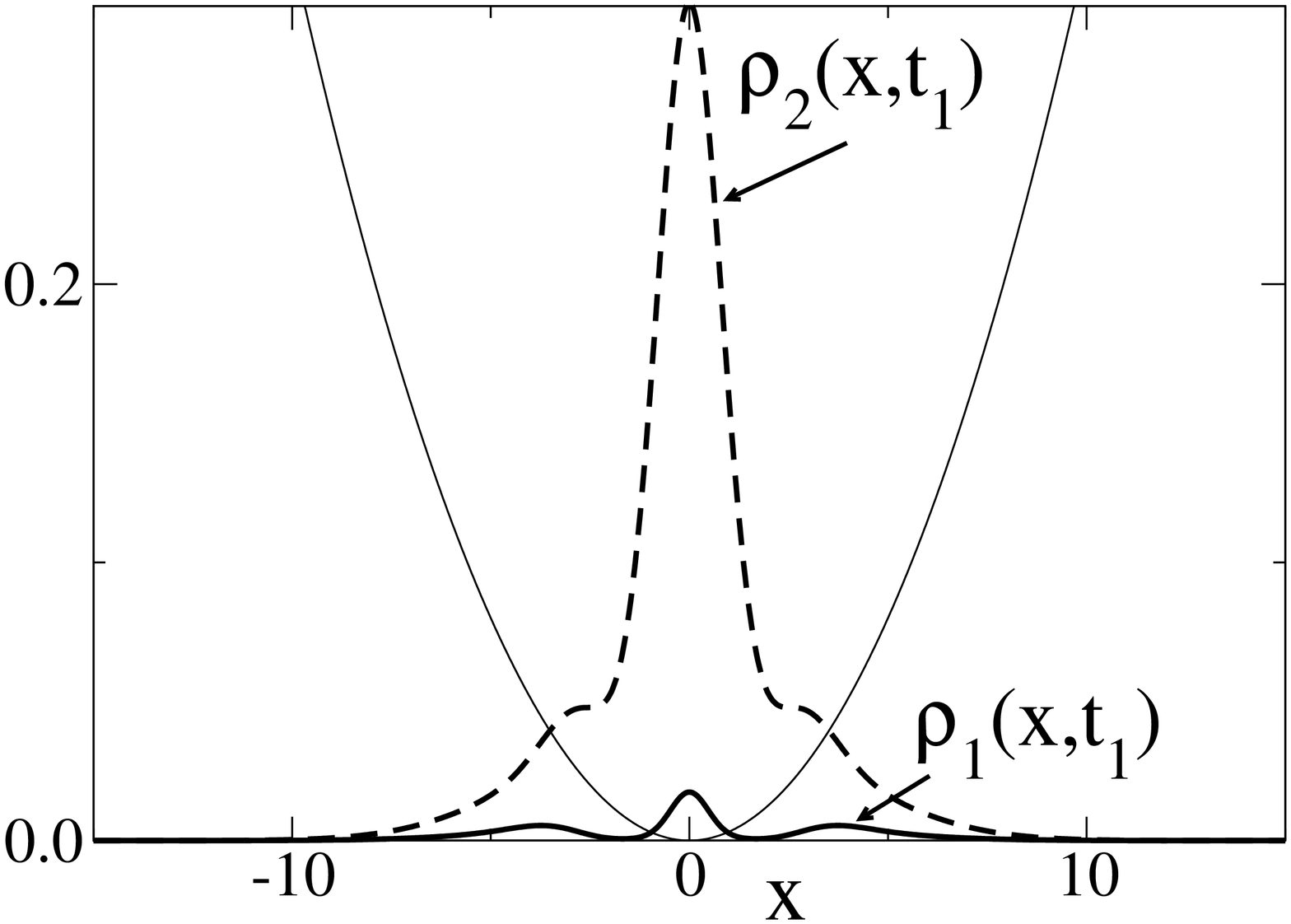} &
\includegraphics[width=4.cm,height=5.cm,angle=270,clip]{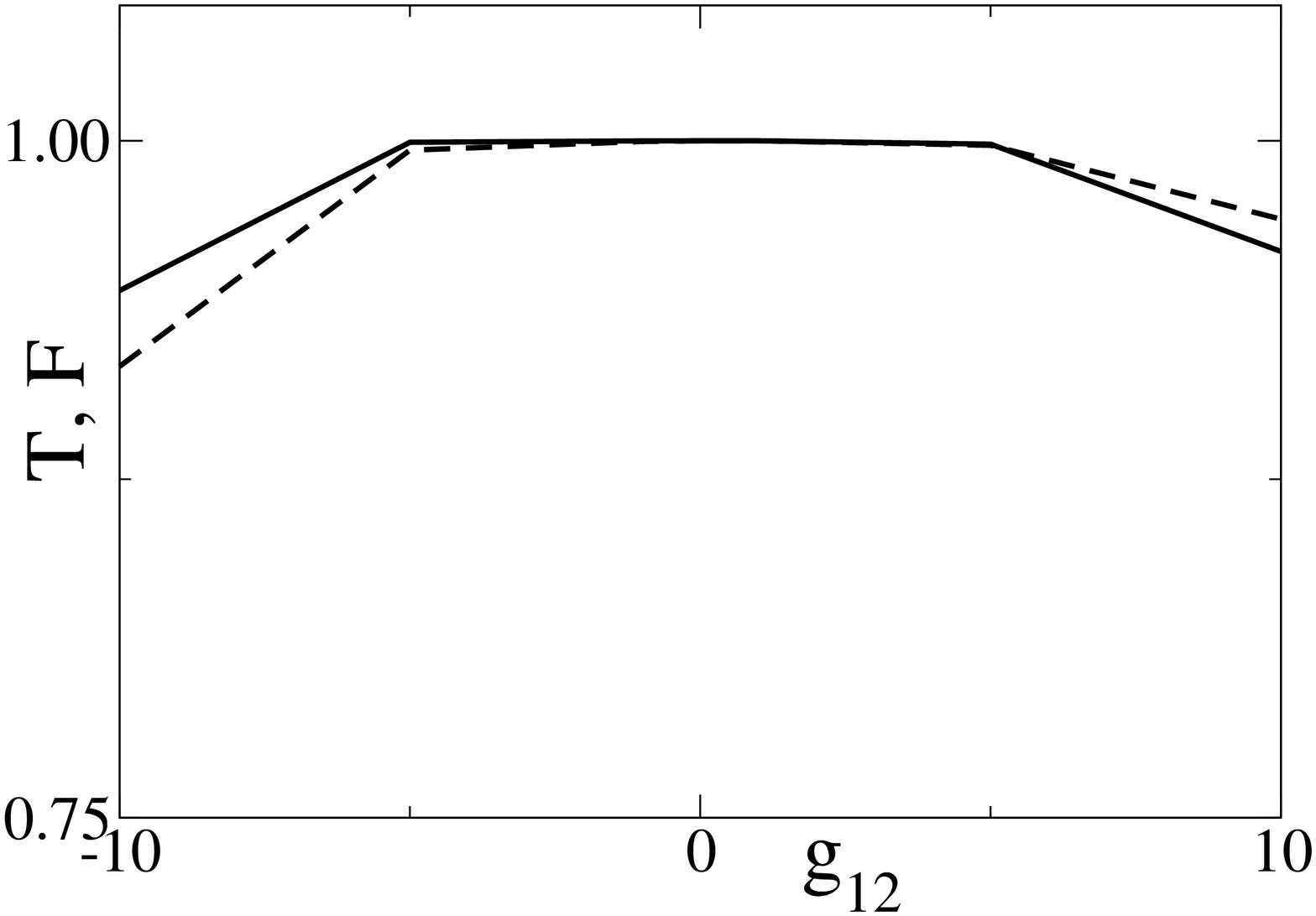}   \\
\end{tabular}
\caption{Panel (a) shows the initial condition: all the particles are in the ground state of the first component. 
The solid line shows the harmonic trapping potential. 
The parameters used in the GP equations are $\Omega=0.08$, $t_0=0$, $\gamma=\pi/10$, $g_{11}=0.97$, $g_{12}=1.03$. 
In (b) we plot the population imbalance $\eta(t)$ for $g_{12}=1$ (solid line), $-1$ (dashed line) and 
$-10$ (dotted line). The optimal times with the maximum transfer are respectively $t_1=5, 5.02, 6.48$. 
In (c), (d) and (e) we show the spatial profile 
$\rho_2(x,t_1)\equiv\vert \psi_2(x,t_1)\vert ^2$ at these optimal times. In (e), where the transfer is not optimal, we plot 
also the the spatial profile $\rho_1(x,t_1) \equiv \vert \psi_1(x,t_1)\vert ^2$. In 
panel (f) we show the transfer efficiency function $T$ (solid line) and the fidelity $F$ (dashed line) vs. the value 
of the inter-species interaction coefficient $g_{12}$.}
\label{rfig2}
\end{center}
\end{figure}

When the $g_{ij}$'s are different, the transfer will no longer be complete. 
As a measure of the deviation from the ``ideal switch'', we characterize the degradation of 
the switch in this inhomogeneous case according to the 
relevant quantities introduced in Eq. (\ref{efficiency}) 
and (\ref{fidelity}). 
For repulsive condensates, we have considered the case of the ground 
state of $^{87}$Rb, where the two spin states mentioned above 
have corresponding scattering lengths such that $g_{11}:g_{22}=0.97:1.03$. 
In this context, we have identified the ground state configuration for the first component alone and have subsequently applied the 
linear coupling for $\gamma=\pi/10$, 
for various values of $g_{12}$. The resulting transfer
efficiency as a function of $g_{12}$ is shown in Fig. \ref{rfig2} (bottom right panel).
The numerical result indicates that in the defocusing regime of repulsive inter-species interactions, 
the transfer efficiency and fidelity remain very high ($>0.9$) throughout the interval $-10<g_{12}<10$. Similar results have been obtained 
for attractive interactions $g_{11}=g_{22}=-1$ and varying $g_{22}$, starting form the ground state of the first component 
in presence of the trap.


\begin{figure}[t]
\begin{center}
\begin{tabular}{cc}
   (a) & (b) \\
\includegraphics[width=4.cm,height=5.cm,angle=270,clip]{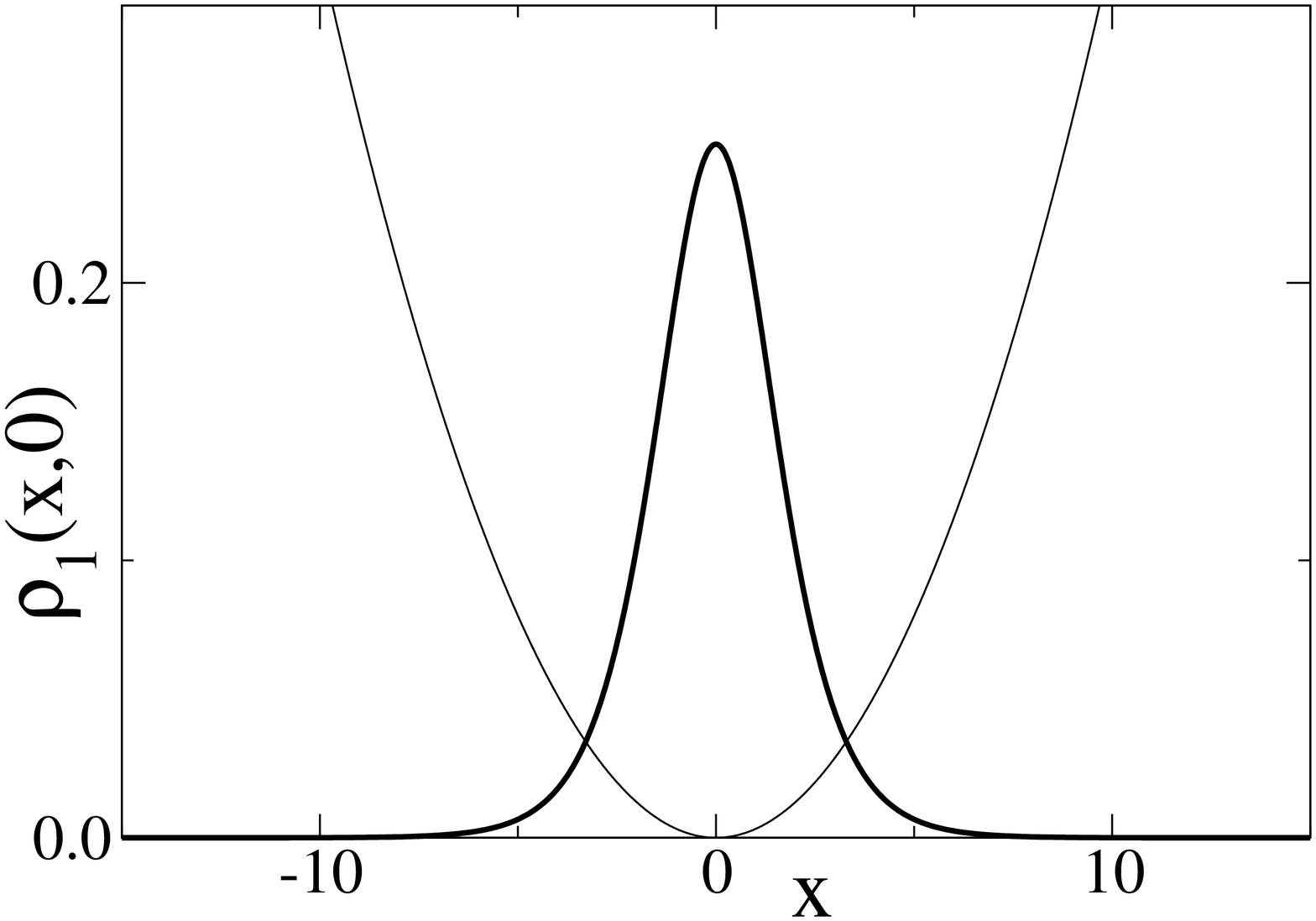} &
\includegraphics[width=4.cm,height=5.cm,angle=270,clip]{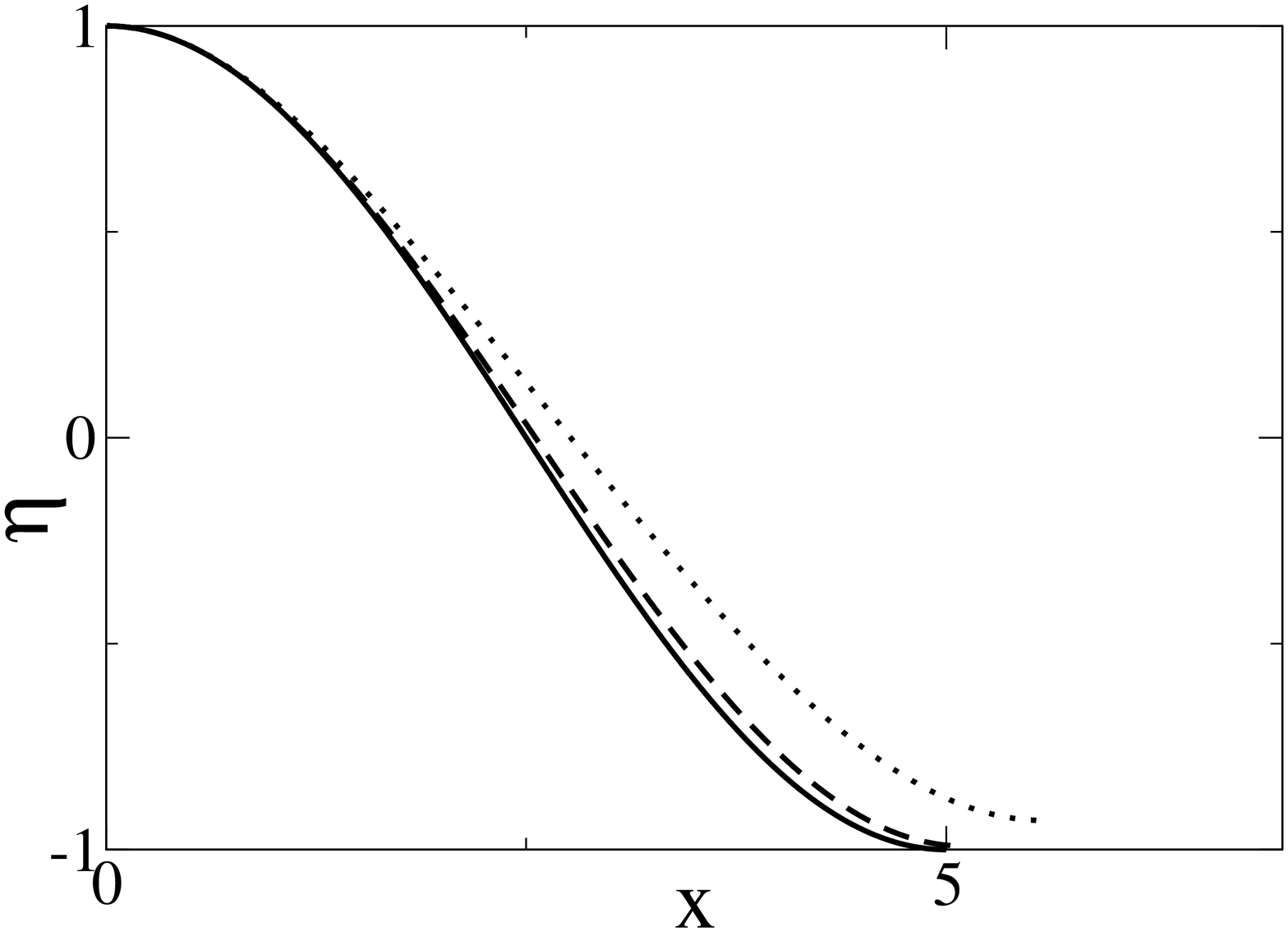} \\
   (c) & (d) \\
\includegraphics[width=4.cm,height=5.cm,angle=270,clip]{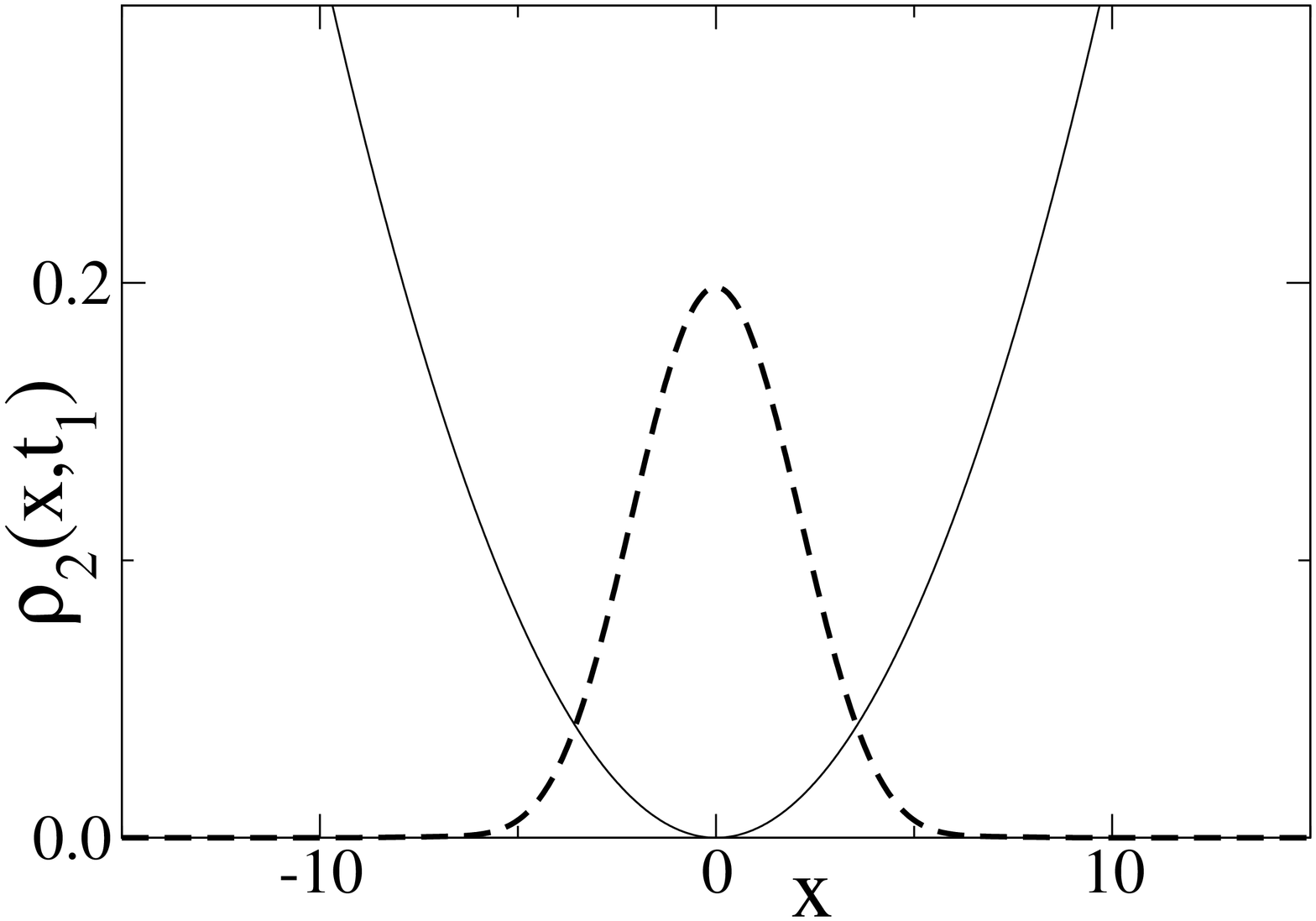} &
\includegraphics[width=4.cm,height=5.cm,angle=270,clip]{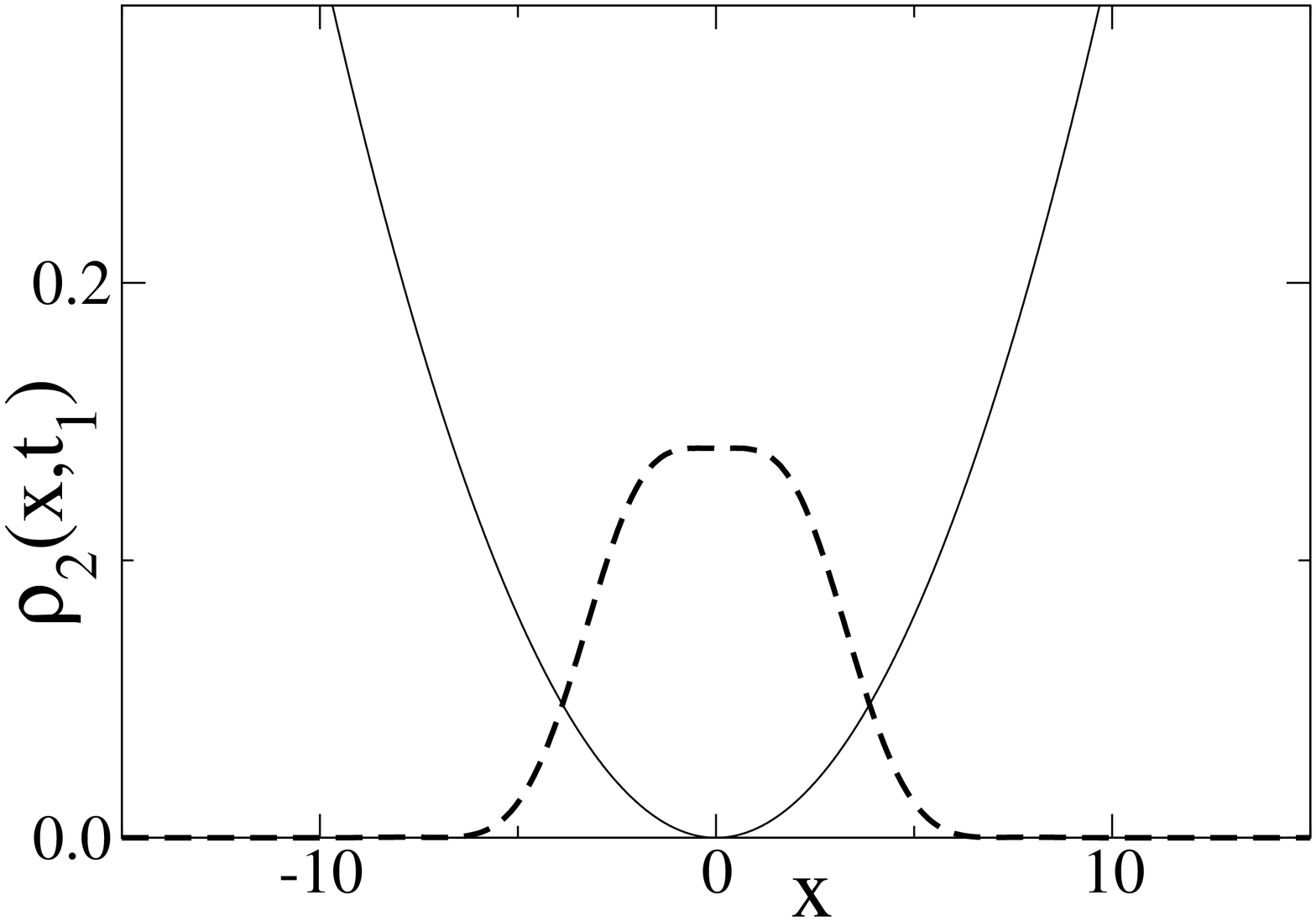} \\
   (e) & (f) \\
\includegraphics[width=4.cm,height=5.cm,angle=270,clip]{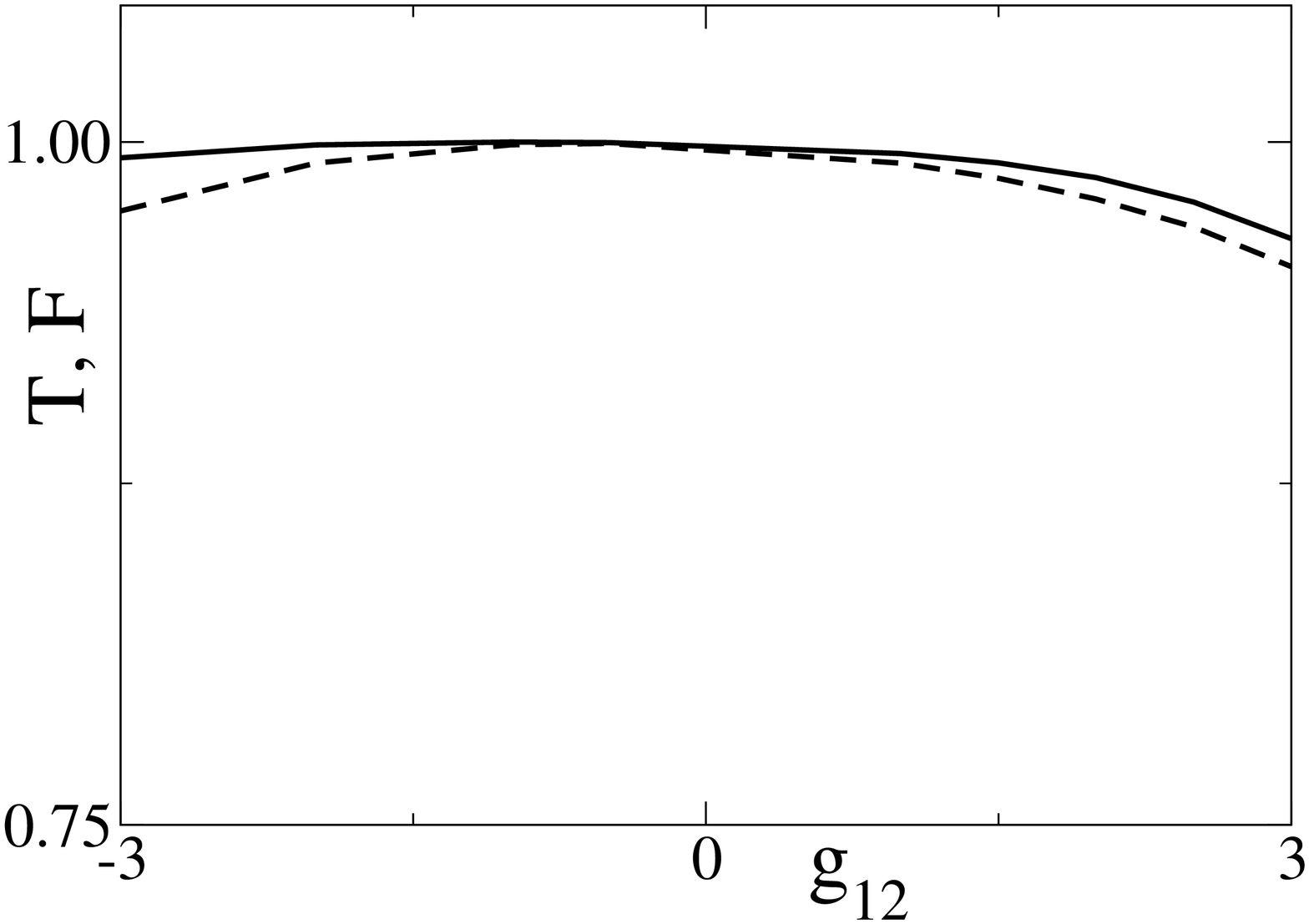} &
\includegraphics[width=4.cm,height=5.cm,angle=270,clip]{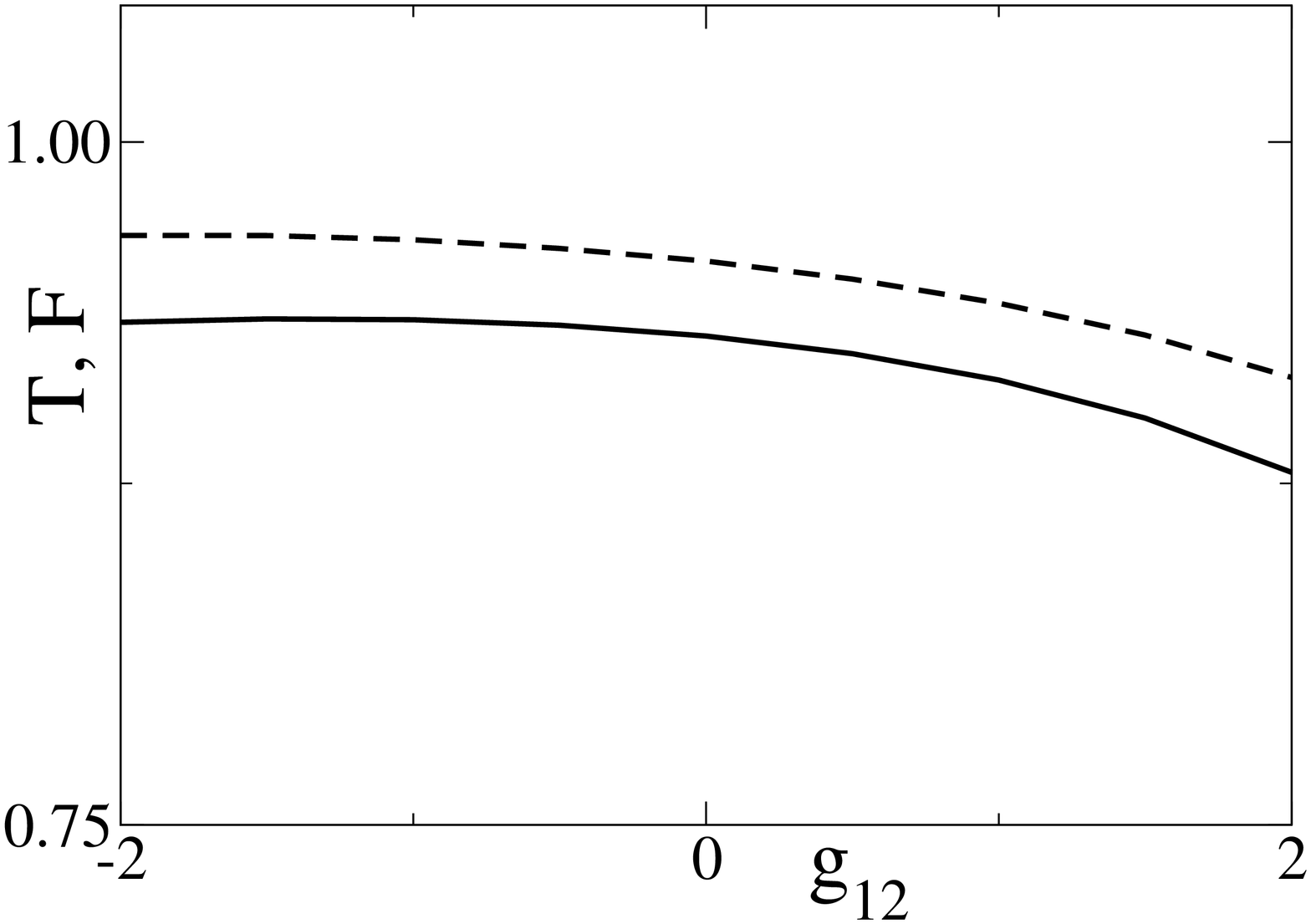} \\
\end{tabular}
\caption{Panel (a) shows the initial condition: all the particles are in the  soliton (\ref{solit}) of the first component. 
The parameters used are $\Omega=0.08$, $t_0=0$, $\gamma=\pi/10$, $g_{11}=g_{12}=-1$. 
In (b) we plot the population imbalance $\eta(t)$ for $g_{12}=-1$ (solid line), $1$ (dashed line) and 
$3$ (dotted line). The optimal times for maximum transfer are respectively $t_1=5, 5.10, 5.58$. 
In (c) and (d) we show the spatial profile of $\rho_2(x,t_1) \equiv
\vert \psi_2(x,t_1)\vert ^2$ at these optimal times 
for $g_{12}=1$ and $g_{12}=3$. In 
panel (e) we show the transfer efficiency function $T$ (solid line) and the fidelity $F$ (dashed line) vs. the value 
of the inter-species interaction coefficient $g_{12}$, and in (f) we plot the same quantities for 
$g_{11}=-1$, $g_{22}=1$ and the same initial condition: the efficiency is reduced with respect to the case 
$g_{11}=g_{22}=-1$.}
\label{rfig3}
\end{center}
\end{figure}

We also performed similar computations in the case of attractive 
intra- and inter- species interactions for the solitonic initial condition (\ref{solit}) in the first component, 
which, with the trap confinement, is no longer the ground state. We have set the intra-species interactions at 
$g_{11}=g_{22}=-1$, varying $g_{12}$. Our findings are plotted in Fig. \ref{rfig3}, showing 
the robustness of our protocol in a range of values of $g_{12}$ between $-3$ and $3$, a range smaller with respect 
to the case in which the initial condition is the ground state of the first component. This is due 
to the fact that the initial condition is not the ground state, and to the fact that 
the Rabi pulse used is homogeneous in space;
note that (according to the analysis of Appendix A) a sort of space-dependent Rabi pulse 
would be needed to improve further the transfer efficiency. 
Finally, as expected, the Rabi switch is less effective in the case in which $g_{11}$ and $g_{22}$ have opposite signs. 
The qualitative reason is 
that the wavefunction, being a bright soliton which is supported by a 
negative coupling constant,  
cannot easily be transferred 
to an environment characterized by a positive coupling constant (which does not support bright solitons but rather dark ones).  
The fidelity and the efficiency are plotted in the panel (f) of 
Fig. \ref{rfig3}.


\section{Results for $2D$ settings}

Let us now consider the $2D$ version of Eqs. (\ref{lceqa})-(\ref{lceqb}), pertaining to 
``pancake''-shaped condensates \cite{reviews}. We will focus on the realistic case of a binary mixture of two 
hyperfine states of $^{87}$Rb, with $g_{11}:g_{12}:g_{22}=0.97:1:1.03$, 
and examine both ground and excited states; the latter, will 
be characterized by the presence of one vortex or of many vortices arranged 
as vortex-lattice configurations. We shall consider only 
repulsive intra-species interactions, since for attractive ones, 
the system is generally subject to collapse \cite{reviews}. 
In order to compare the results with the ones 
pertaining to the $1D$ setting, 
we vary the inter-species strength $g_{12}$ in the same domain  
(although for $g_{12}<0$ the system is also subject to collapse). 
In a real experiment this may 
be done by using external magnetic fields, which can change the 
magnitude and sign of 
the scattering length through the Feshbach resonance mechanism \cite{books}.

\begin{figure}[t]
\includegraphics[width=8.cm,height=5.cm,angle=0,clip]{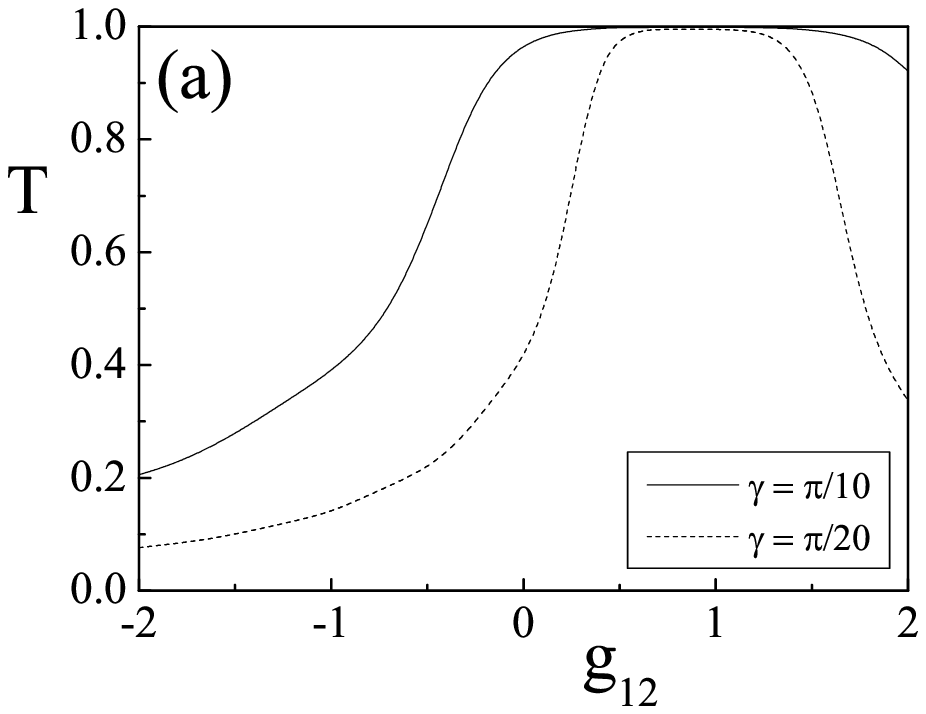}
\\
\includegraphics[width=4.cm,height=5.cm,angle=0,clip]{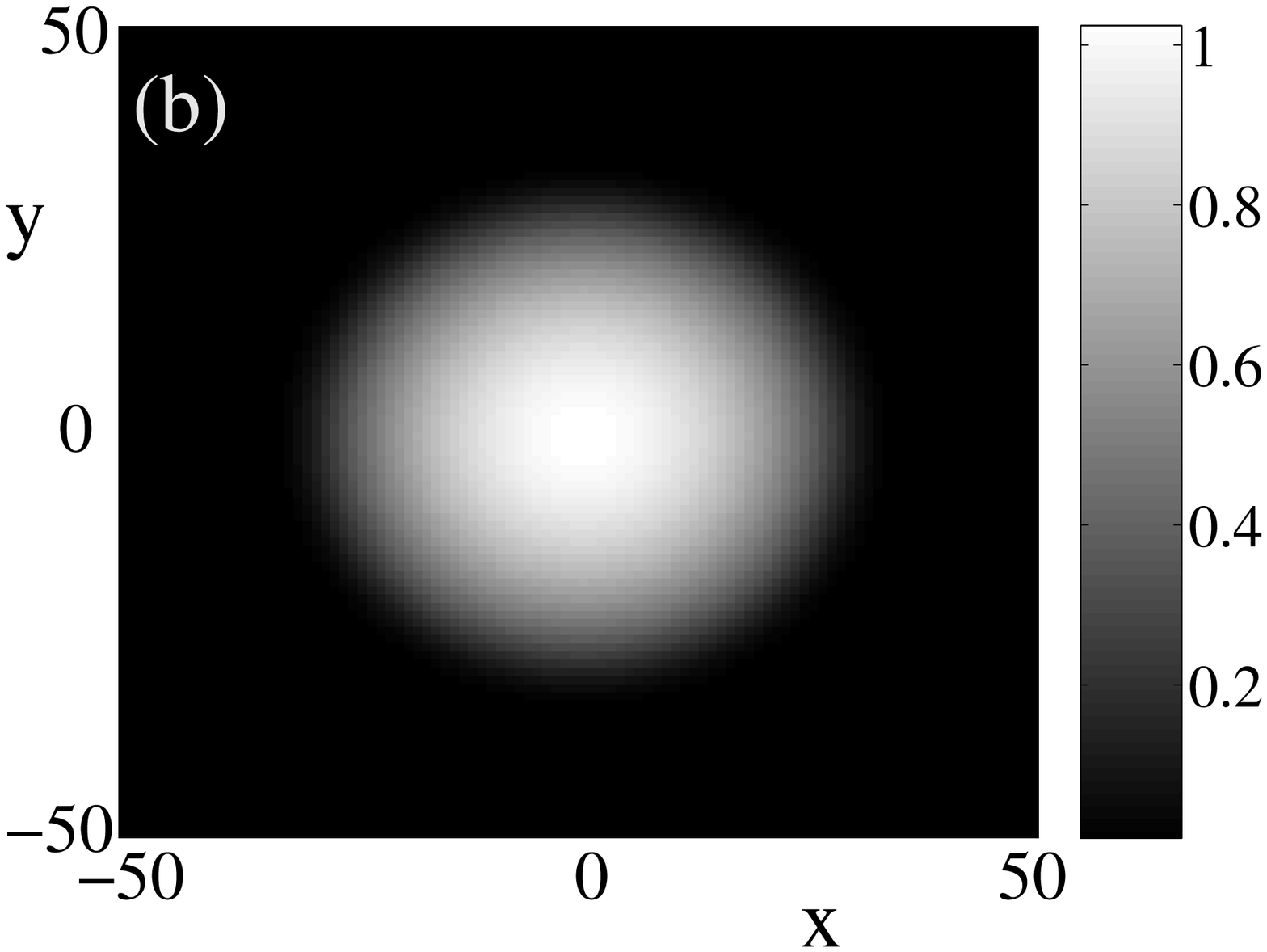}
\includegraphics[width=4.cm,height=5.cm,angle=0,clip]{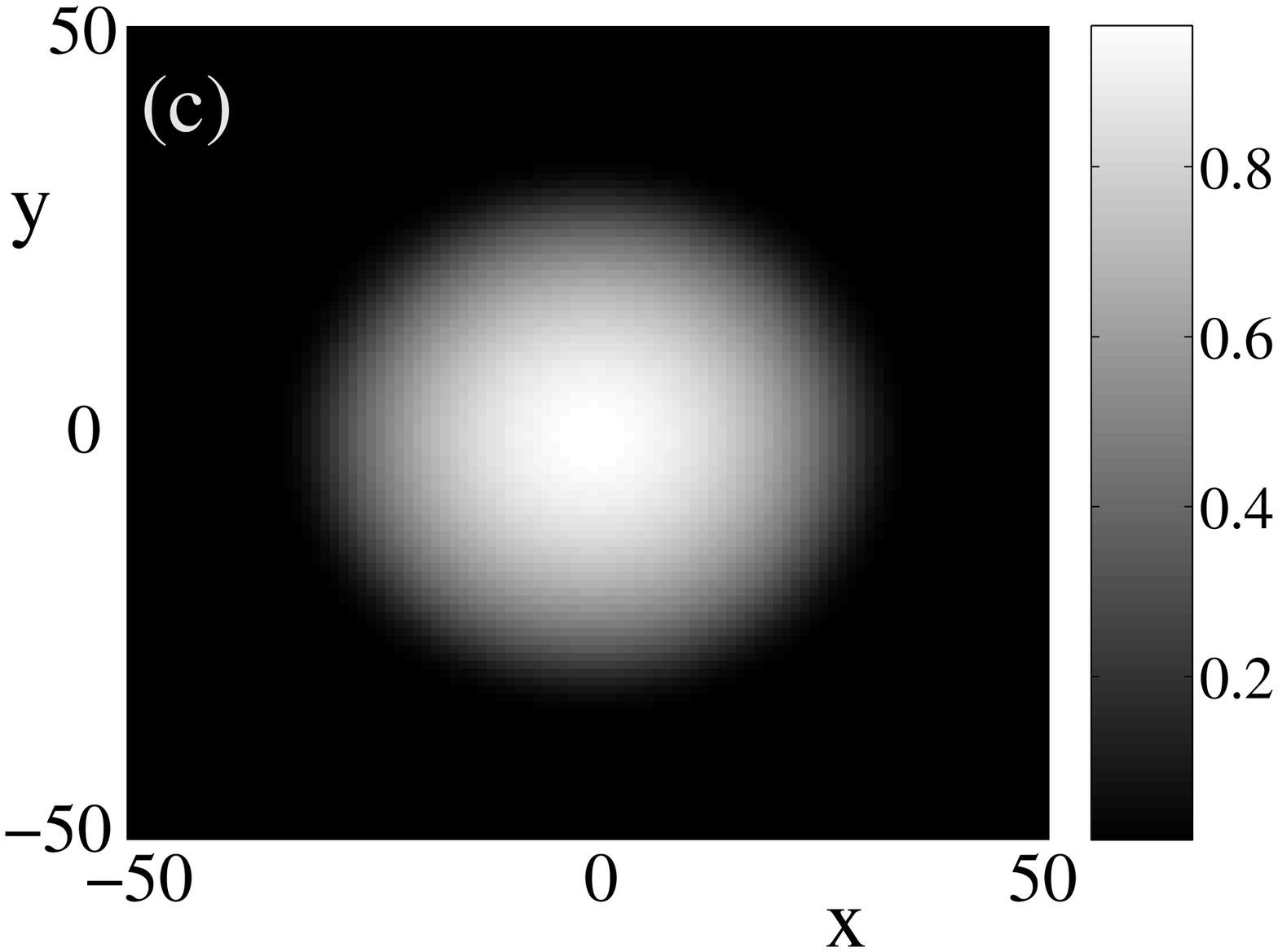}
\\
\includegraphics[width=4.cm,height=5.cm,angle=0,clip]{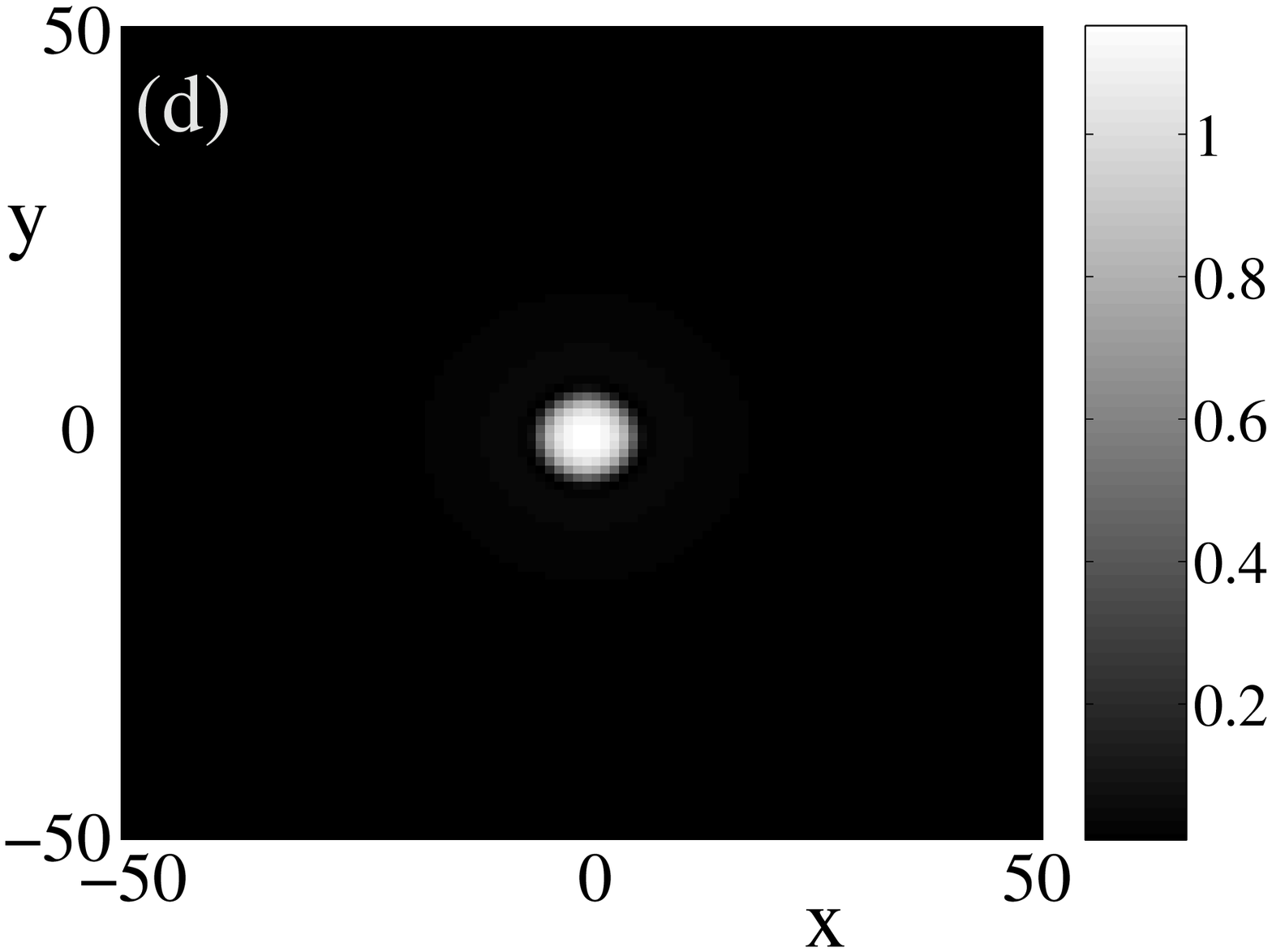}
\includegraphics[width=4.cm,height=5.cm,angle=0,clip]{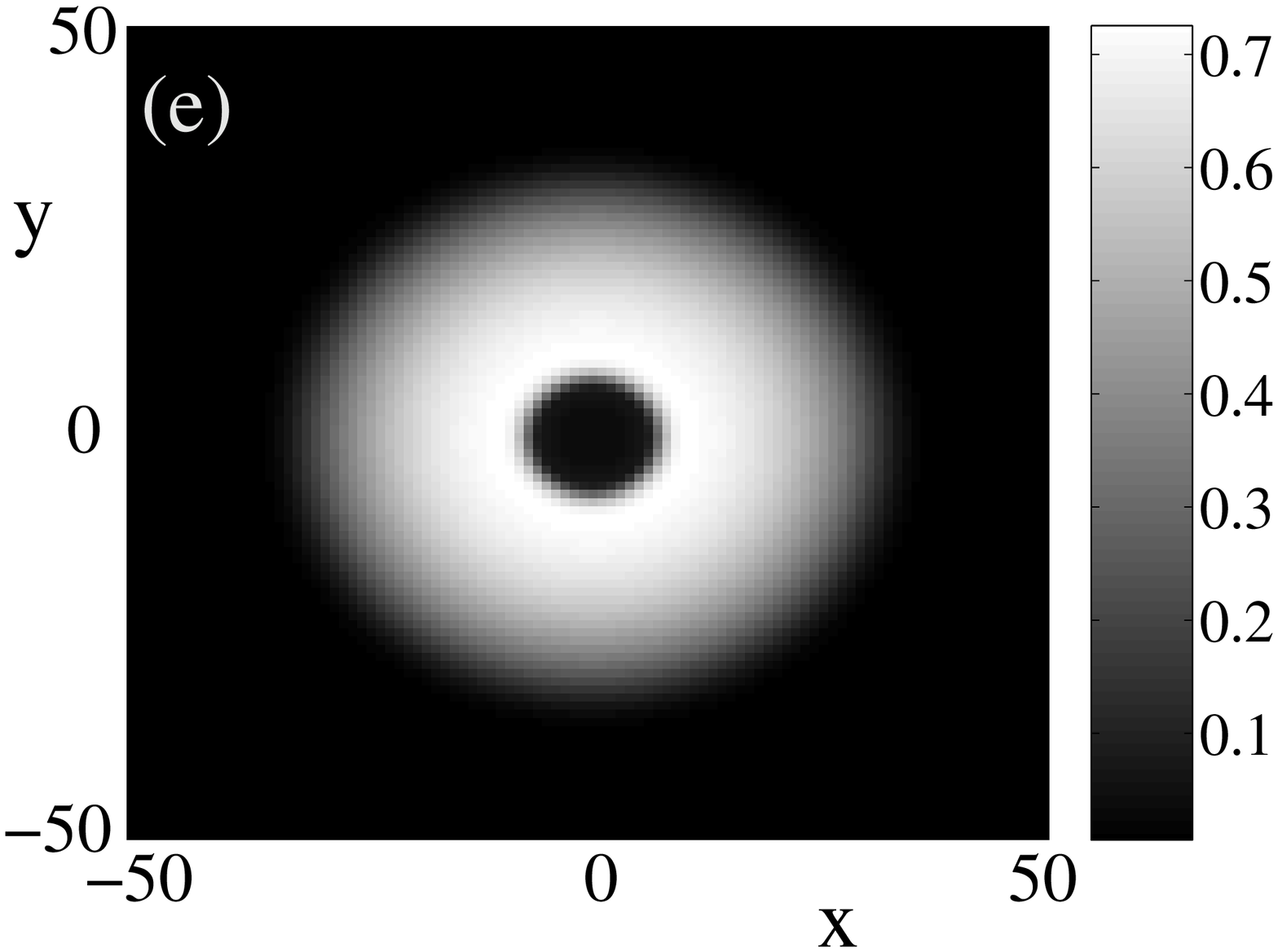}
\caption{(a) The transmitivity function $T$ versus the inter-species strength $g_{12}$ for $\gamma=\pi/10$ (solid line) 
and $\gamma=\pi/20$ (dashed line).
(b) Contour plot of the initial density of the ground state in 
the first component $\psi_1$. (c) Contour plot of the corresponding final density $\vert \psi_{2}(t=25)\vert ^2$ 
of the second component for $g_{12}=1$; here the transfer of matter is complete. 
(d), (e) Contour plots of the corresponding final densities of both species, namely $\vert \psi_{1}(t=25)\vert ^2$ 
and $\vert \psi_{2}(t=25)\vert ^2$, but 
for $g_{12}=2$; as it is seen part of the matter remains (is missing) in the first (from the second) component.}
\label{tf2d}
\end{figure}

It is relevant to consider at first the efficiency $T$ 
as a function of $g_{12}$, for different values of $\gamma$. The result is shown in the top panel of 
Fig. \ref{tf2d}. In the 2D case, the transfer is almost complete 
as $T \ge 0.95$ for all positive values of $g_{12}<2$ (for $\gamma=\pi/10$). 
It is also seen that the efficiency is higher for larger values of 
the magnitude (or inverse duration) of the linear coupling coefficient $\gamma$.

We have considered the ground state of the system, shown in Fig. \ref{tf2d}(b), in the first 
component $\psi_1$, for an harmonic trapping potential with strength $\Omega=0.045$ (the chemical 
potential is equal to one). In this case, assuming that the switch parameters are $t_0=10$ and $\gamma=\pi/10$,  
we have found the following: for $g_{12}=1$, the transfer of matter in the second component $\psi_2$ is complete
[see Fig. \ref{tf2d}(c) where the density of the second component at $t=25$ is shown], while for $g_{12}=2$ it is 
incomplete. In particular, as shown in Figs. \ref{tf2d}(d)-(e) (where the densities 
$\vert \psi_{1}(t=25)\vert ^2$ and $\vert \psi_{2}(t=25)\vert ^2$ are respectively shown), 
a fraction of matter remains in the first component and is correspondingly missing from the central part of the 
second component after the switch-off of the Rabi pulse.

\begin{figure}[t]
\includegraphics[width=8.cm,height=5.cm,angle=0,clip]{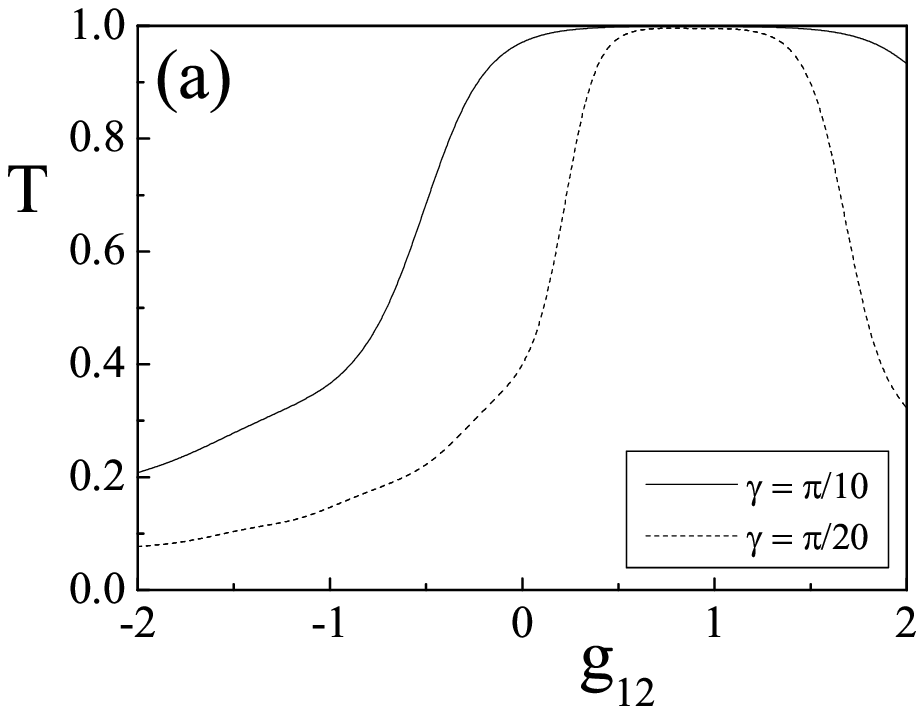}
\\
\includegraphics[width=4.cm,height=5.cm,angle=0,clip]{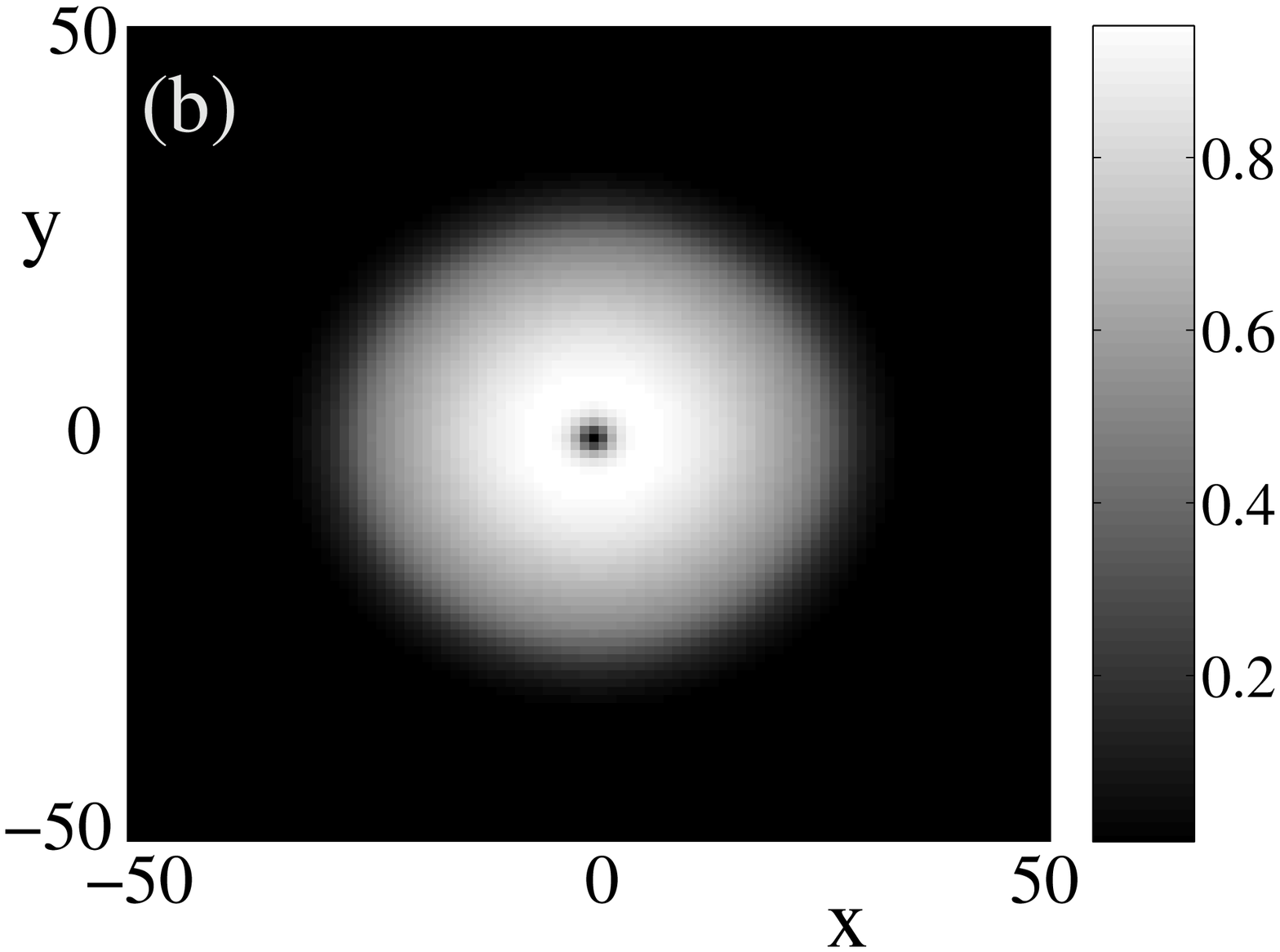}
\includegraphics[width=4.cm,height=5.cm,angle=0,clip]{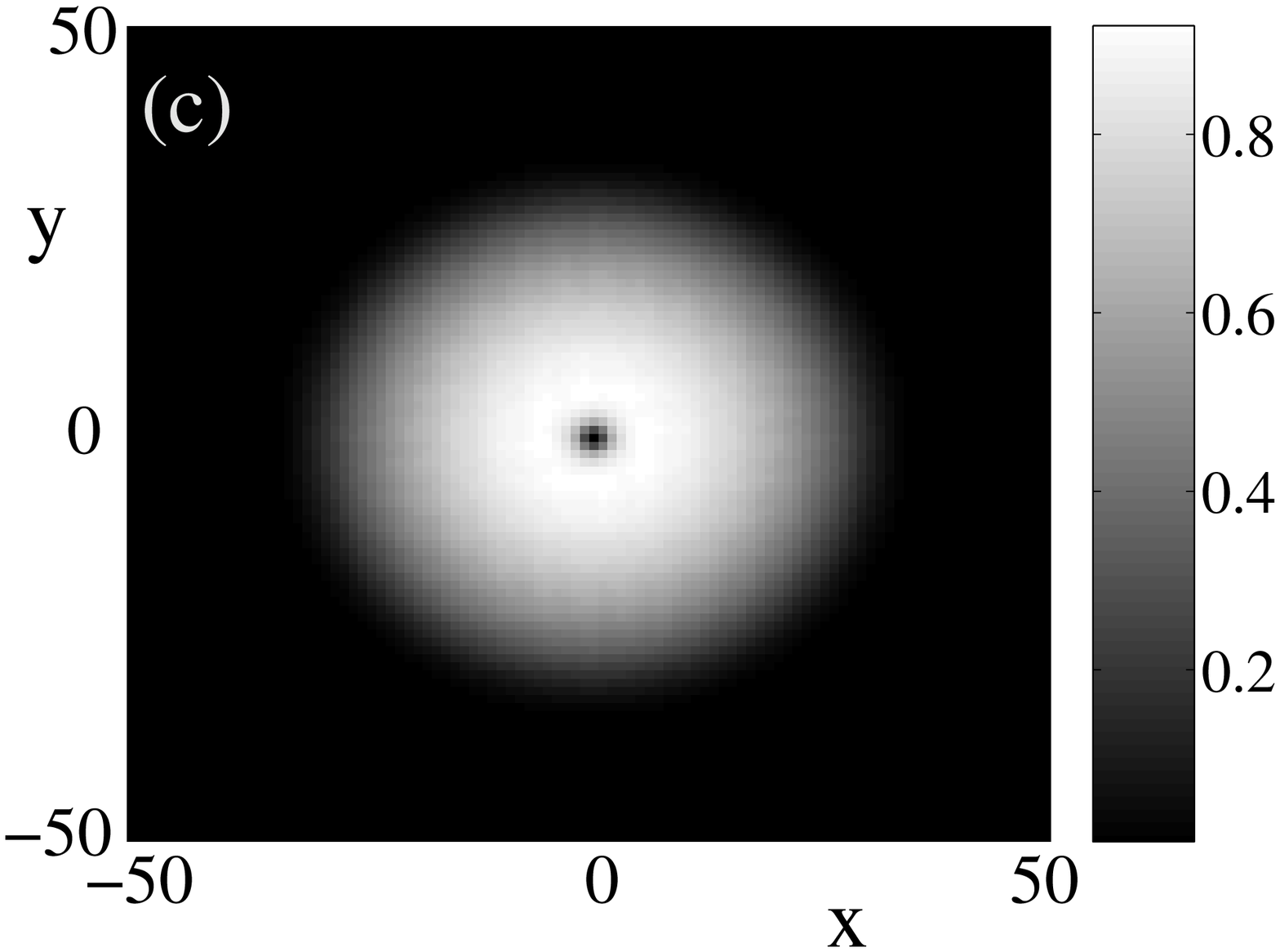}
\\
\includegraphics[width=4.cm,height=5.cm,angle=0,clip]{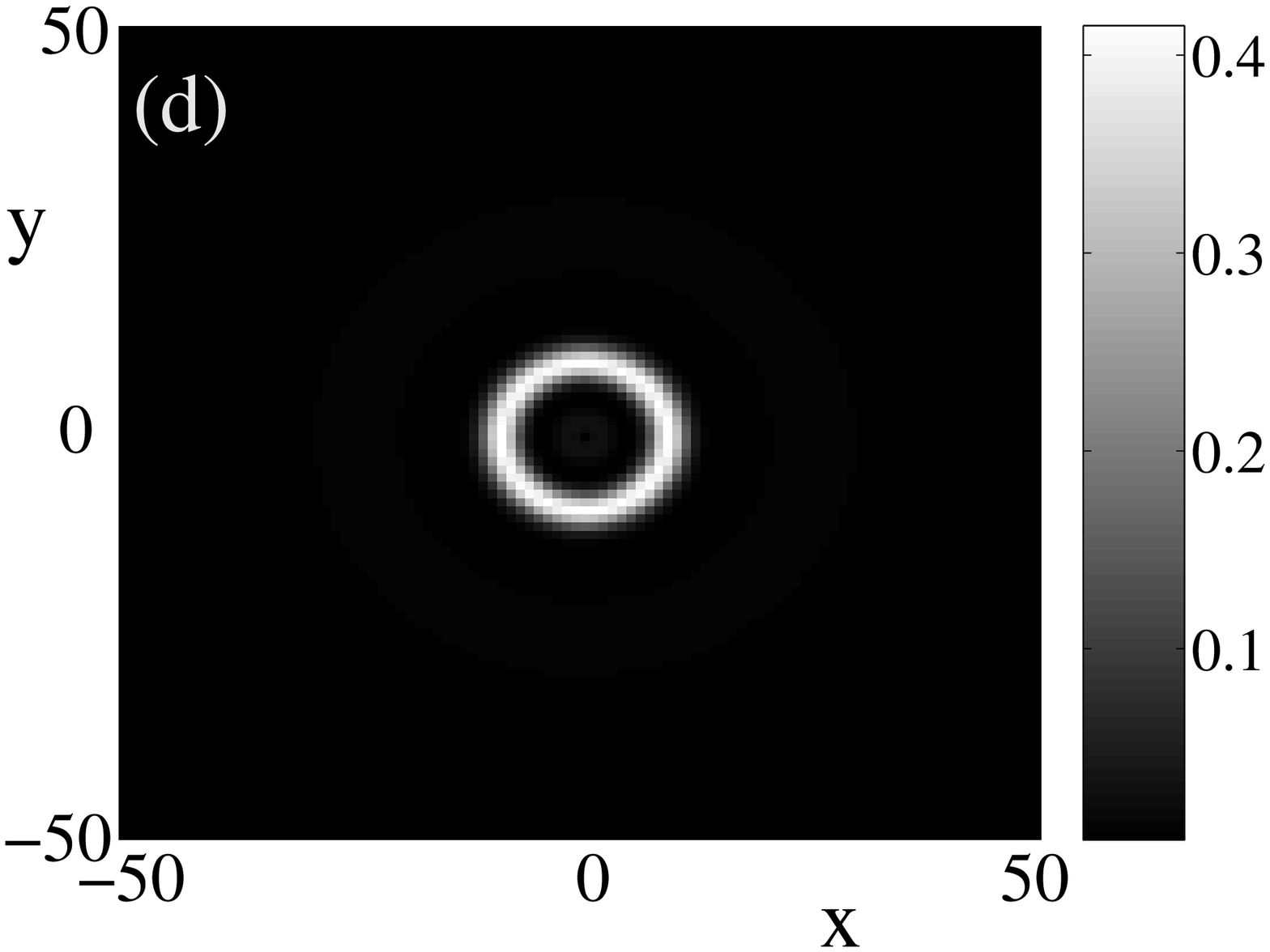}
\includegraphics[width=4.cm,height=5.cm,angle=0,clip]{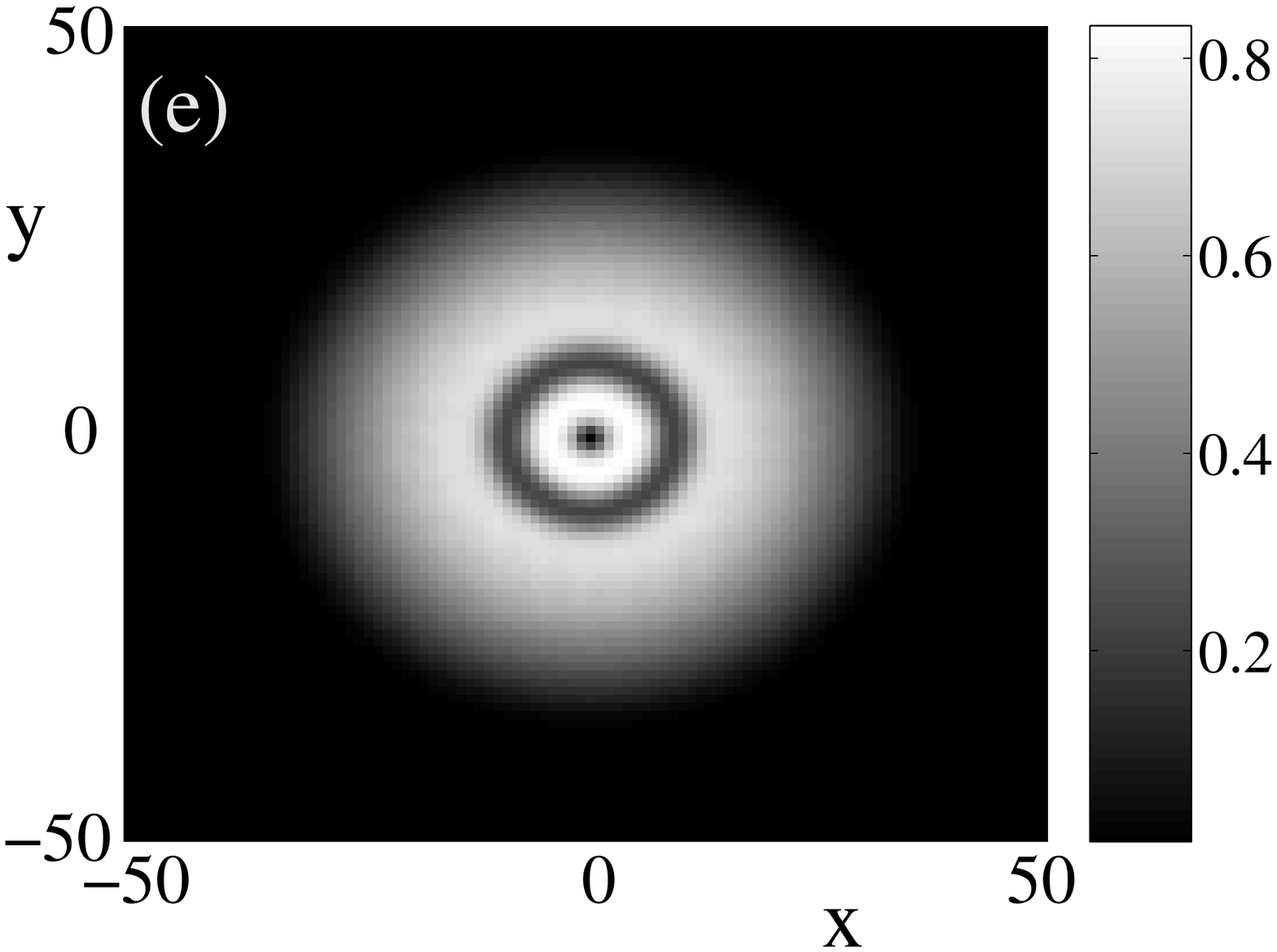}
\caption{(a) Same as in Fig. \ref{tf2d}(a).
(b) The initial density of the first component $\psi_1$, consisting of a 
cloud with one vortex in the center. 
(c) The corresponding final density $\vert \psi_{2}(t=25)\vert ^2$ 
of the second component for $g_{12}=1$; here the transfer is complete and the final configuration is identical to the 
initial one. (d), (e) The corresponding final densities of both species, namely $\vert \psi_{1}(t=25)\vert ^2$ and 
$\vert \psi_{2}(t=25)\vert ^2$, but for $g_{12}=2$.}
\label{vo2d}
\end{figure}

Next we consider an excited state, in which a vortex is initially placed at the center of the BEC cloud
(first component). Here, it is interesting to investigate whether such a coherent nonlinear state can be transfered 
in the second component. As seen in Fig. \ref{vo2d}(a), 
the efficiency is as high as for the ground state transfer. 
Also, for $g_{12}=1$, a perfect transfer of this excited state occurs as seen 
in Figs. \ref{vo2d}(b) and (c), where the initial state of the first species and the final one of the second 
species are respectively shown. 
Nevertheless, for $g_{12}=2$, the transfer is not complete, as it can be seen 
in Figs. \ref{vo2d}(d) and (e): starting again from the initial density shown in in Fig. \ref{vo2d}(b), 
after the switch-off of the process, a ring-shaped 
part of the matter remains in (is missing from) the first (second) component. 
A careful observation of Fig. \ref{vo2d}(d) also 
shows that this ``high'' density ring surrounds a low density ($\approx 0.2$) part of matter with a vortex on top of it. 
Note that even in this case, the vortex is transferred in the second component; on the other hand, the above mentioned 
``bright'' and ``dark'' ring structures (respectively in the first
and second component) do not carry any 
topological charge. It is interesting to remark that such methods 
are similar in spirit to the ones used to produce ring-like patterns
in the recent experimental work of \cite{mertes07}.


\begin{figure}[t]
\includegraphics[width=8.cm,height=5.cm,angle=0,clip]{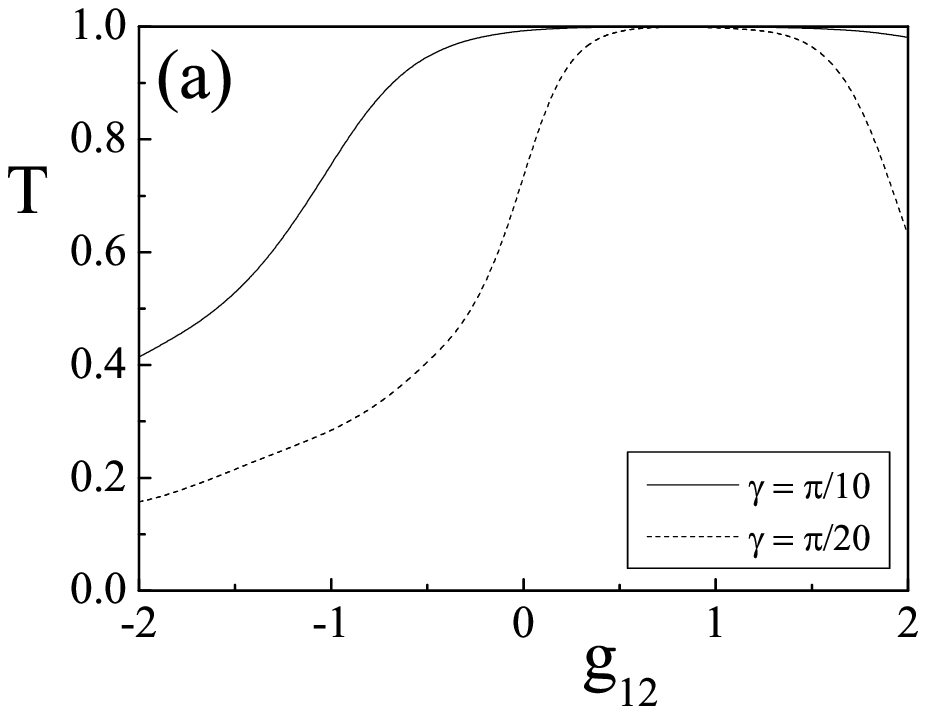}
\\
\includegraphics[width=4.cm,height=5.cm,angle=0,clip]{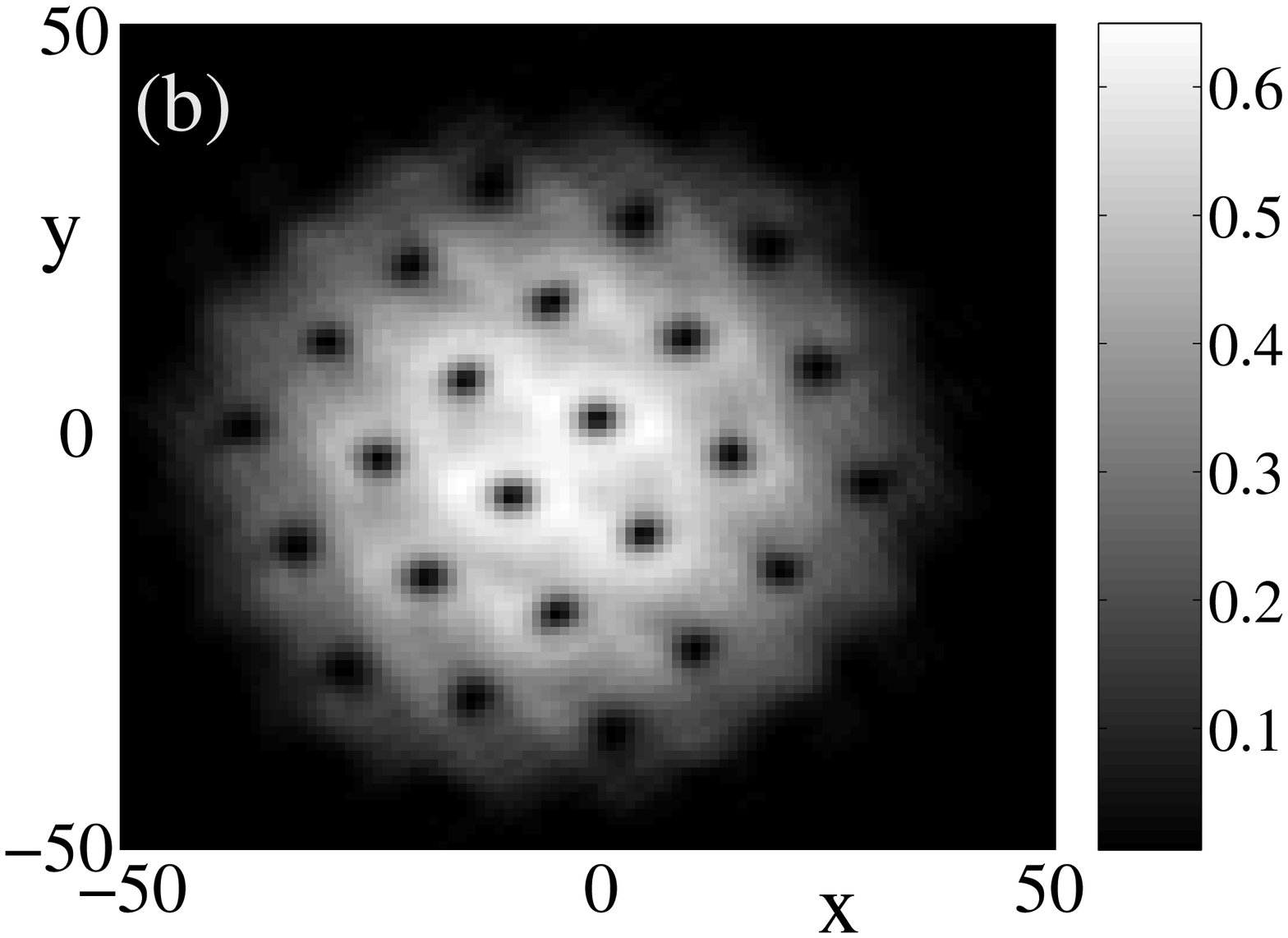}
\includegraphics[width=4.cm,height=5.cm,angle=0,clip]{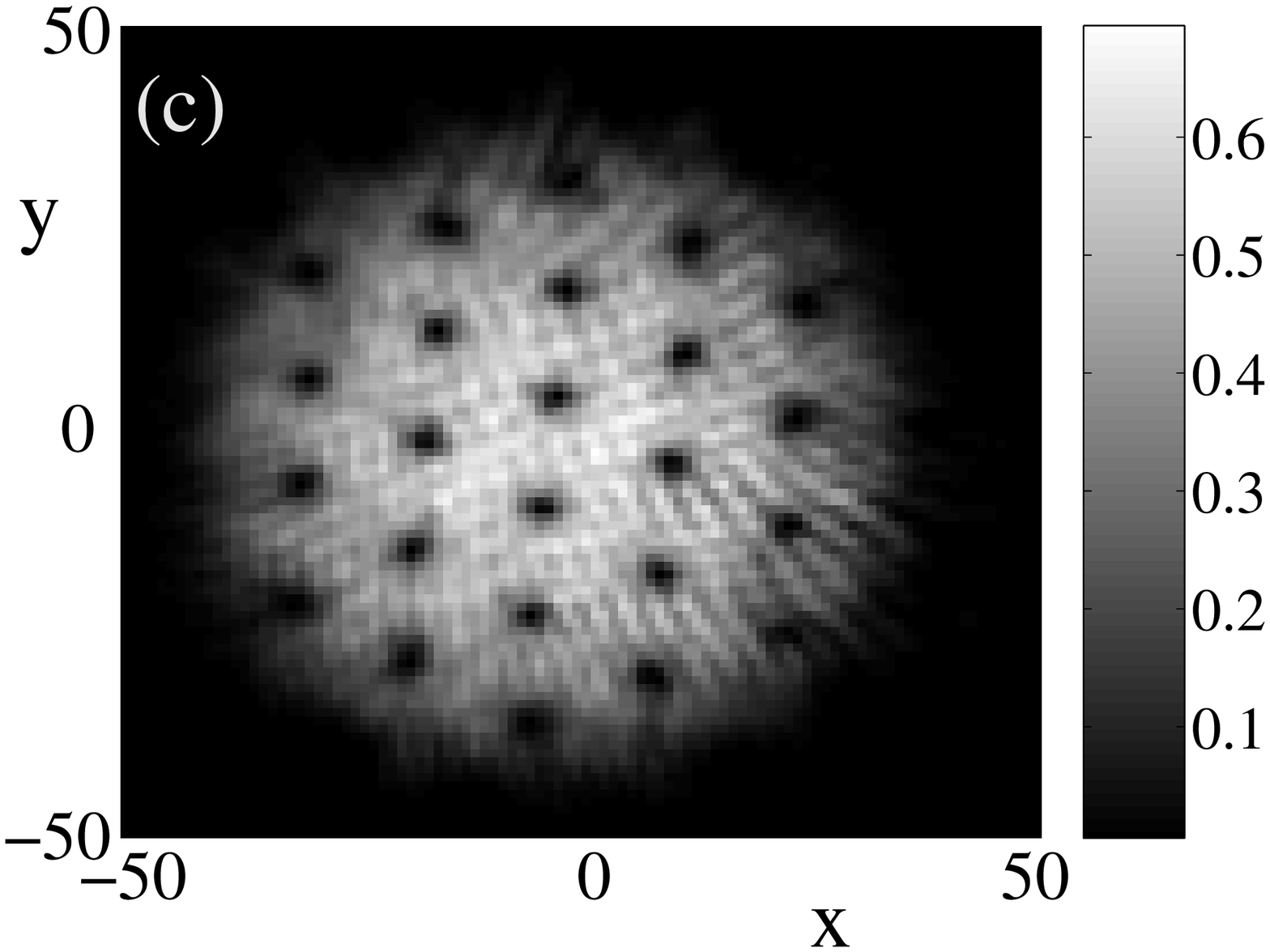}
\\
\includegraphics[width=4.cm,height=5.cm,angle=0,clip]{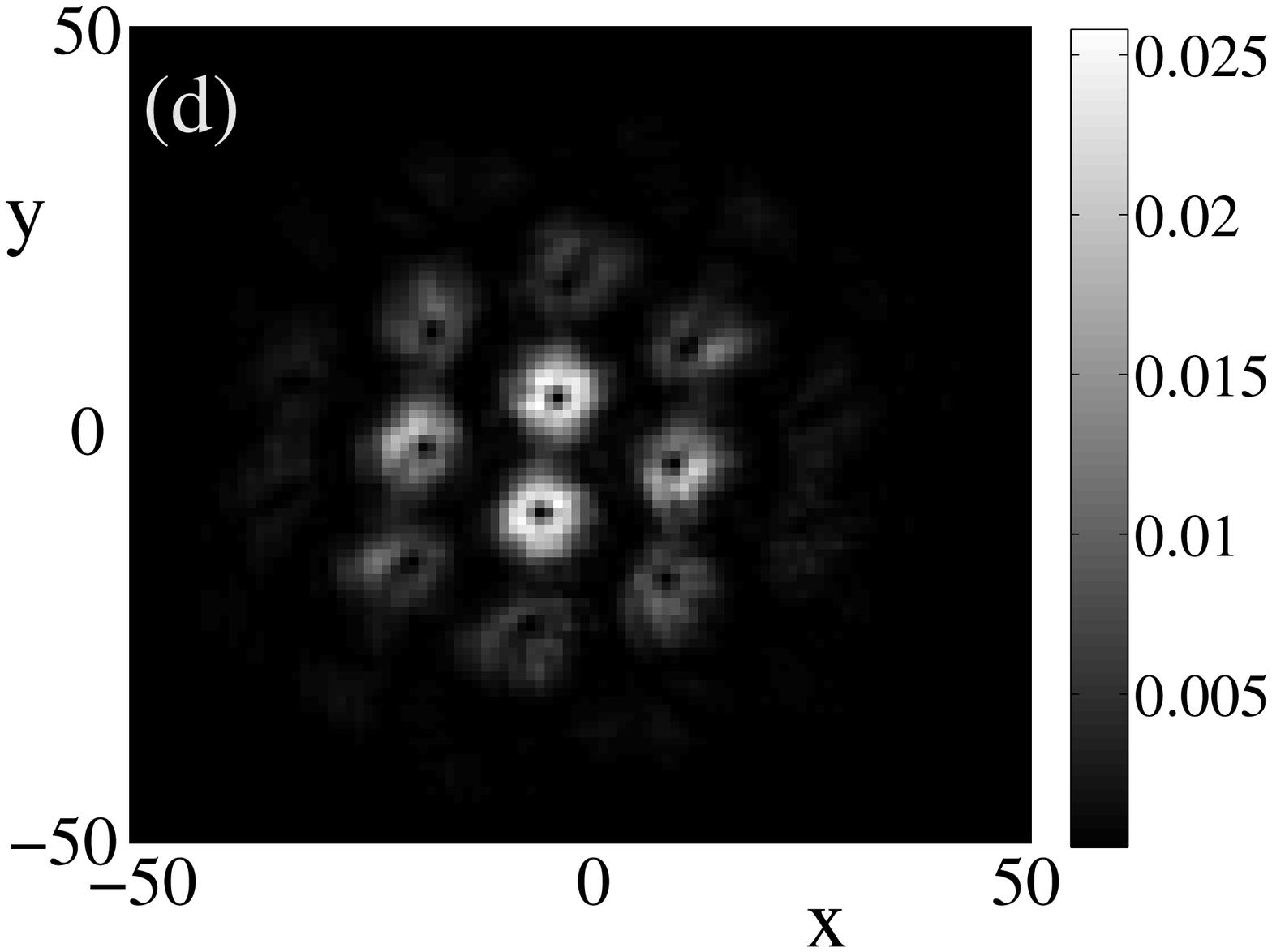}
\includegraphics[width=4.cm,height=5.cm,angle=0,clip]{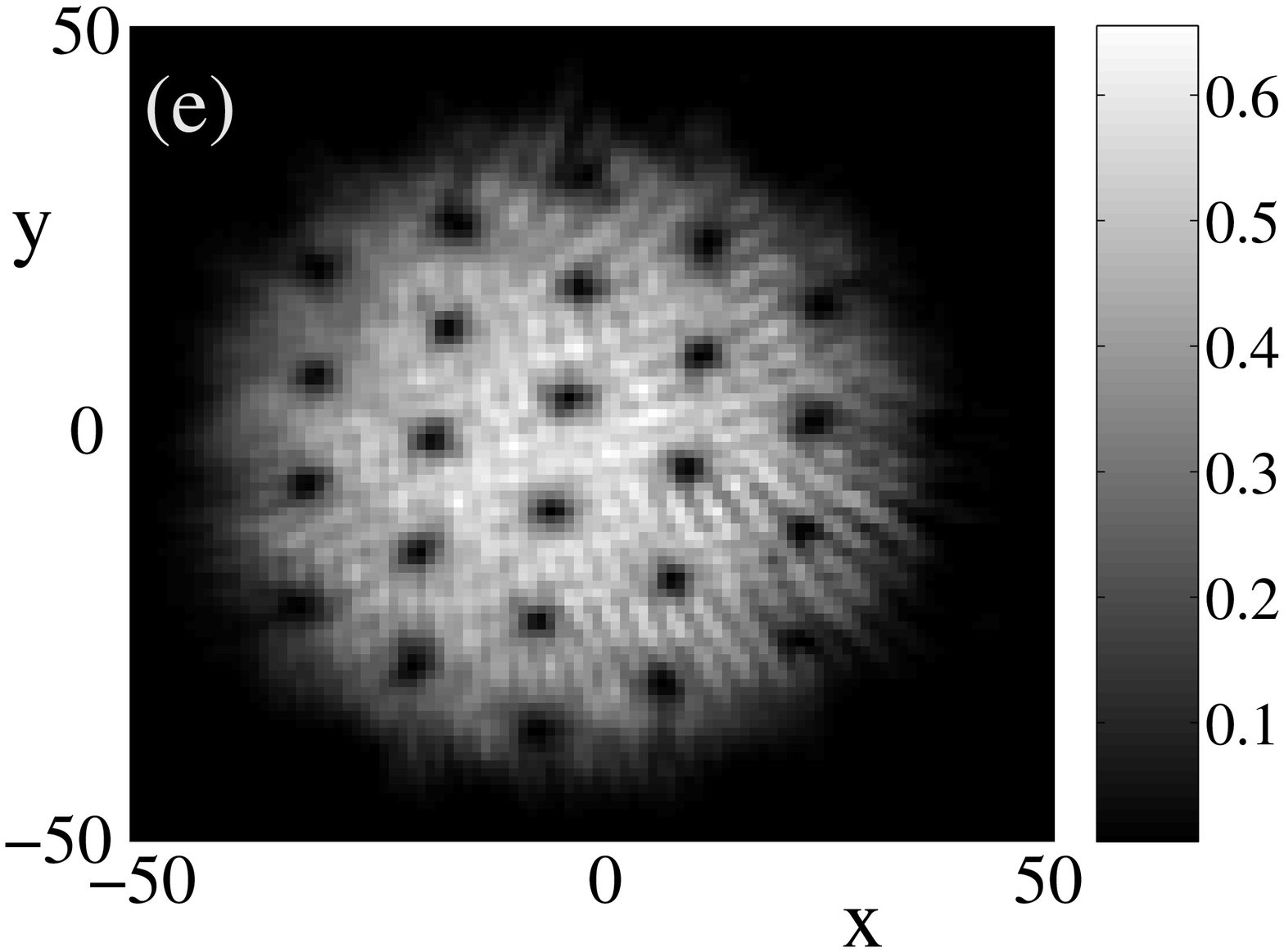}
\caption{(a) Same as in Figs. \ref{tf2d}(a) and \ref{vo2d}(a).
(b) The initial density of the first component $\psi_1$, consisting of a TF-cloud with a triangular vortex lattice 
($\approx 24$ vortices). (c) The corresponding final density $\vert \psi_{2}(t=25)\vert ^2$ 
of the second component for $g_{12}=1$; here the TF cloud and the vortex lattice are perfectly transferred to 
the second species. 
(d), (e) The corresponding final densities of both species, 
namely $\vert \psi_{1}(t=25)\vert ^2$ and $\vert \psi_{2}(t=25)\vert ^2$, but 
for $g_{12}=2$.}
\label{vl2d}
\end{figure}

Finally, we have considered a vortex cluster, 
namely a triangular vortex lattice, initially 
placed on top of the BEC of the first component. Similarly to what was found 
for the ground state and the single vortex, 
we find that the transfer efficiency function shown in Fig. \ref{vl2d}(a) 
assumes values very close to $1$
for a wide range of values of $g_{12}$ (especially so in the case of short
pulse durations i.e., fast transfer). 
As shown in the example of Figs. \ref{vl2d}(b) and (c) for $g_{12}=1$ 
the vortex lattice in the first 
component is perfectly transferred to the second one. 
An ``imperfect transfer'', for $g_{12}=2$, is  
shown in Figs. \ref{vl2d}(d) and (e), corresponding to the final states of the first and second species 
when the initial state is as in Figs. \ref{vl2d}(b). Here it is observed 
that  
small spots of matter (with densities $\approx 0.025$), each of which 
carries a vortex on it, 
remain in the first species, while a robust crystal structure is 
transferred to the second component. It is worth noting here that the
vortex lattice remarkably 
appears to be even more robust, in its ``switching properties'', 
than the ground state of the 2D system.


\section{Conclusions and outlook}

In this paper, we have systematically 
examined the possibility of using
the linear Rabi coupling between the two (or more) components of a 
Bose-Einstein condensate 
as a means of controllably transferring the wavefunction of one condensate 
to the other. 
In particular, we have focused
on the case of different hyperfine states, even though similar considerations
are applicable to 
multicomponent condensates composed by different atomic species. We have
illustrated that this transfer is exact and can be analytically 
studied in the limit where all inter- and intra- species interactions
are equal. In addition, we have studied departures from this limit 
both numerically and by means of 
a two-mode ansatz, showing that in this two-mode description 
the impossibility to transfer all the particles from a condensate to the other 
is seen as the self-trapping of the initial condensate wavefunction. 
The threshold for the  
self-trapping has been compared in the homogeneous limit with the findings 
of the numerical simulations Gross-Pitaevskii equations. 

The two-mode analysis shows that that for deviations 
from the ideal case one can optimize the 
transfer by choosing 
an appropriate pulse duration different from the duration 
given by Eq. (\ref{pulse}) for equal interaction strengths . 
In the presence of external trapping potentials, our numerical 
simulations show that, for repulsive condensates (but, to
a lesser extent, also for attractive condensates), the Rabi switch 
is very robust with high efficiency. In general, 
a change of sign of the inter-species interactions in $1$D has been shown 
to degrade --although, under appropriate conditions, not substantially-- 
the efficiency of the transfer process. 
The switching was surprisingly found to be even more robust 
for particular types of coherent patterns, such as vortex lattices in 
pancake-shaped condensates. Furthermore, we have also illustrated that
the generalization of the proposed Rabi switch to more than $2$ components can 
offer a possibility for systematically routing matter, in a controllable way, between different ``atomic channels''. 

It would be quite challenging 
to examine experimental realizations of such atomic 
switches and routers especially within the context of two hyperfine states, but also in that of spin-$1$ 
states recently studied in \cite{higbie05,sadler06}. Of particular interest in this setting would be the dynamics and phase
separation of vortices and vortex lattices in 
multicomponent condensates. We expect that our 
mechanism may be relevant when transferring a condensate wavefunction also 
in different components of a dipolar BEC \cite{fattori07}.

Another relevant issue concerns the analysis of the 
role of the quantum fluctuations 
on the Rabi switch of wavefunctions: 
indeed, for two condensates 
(whose dynamics is described by coupled Gross-Pitaevskii equations) 
the Rabi switch is not copying the full many-body wavefunction of a 
component in the other, but only the (one-body) order parameter. 
Then, we expect that the presence of quantum fluctuations 
would naturally degrade the efficiency of our protocol. 
A study of such degradation is relevant for the implementation 
of a quantum register through a two-component Bose gas 
in an array of double-wells. 

The protocol proposed in this paper allows the possibility 
to copy the wavefunction of a condensate into another 
species, providing a matter-wave 
counterpart for optical switches realized in nonlinear fiber optics \cite{agrawal}. ``All-optical switches'' are very important tools 
in this context, allowing for to manipulate and eventually store the information contained in optical solitons. For this reason,  
the study of protocols focused on copying of wavefunctions in Bose 
condensates is deemed to be an important task towards 
an efficient manipulation of matter-wave solitons. Furthermore, one can think 
of a possible application 
of the present protocol to implement copies for quantum registers: indeed, one could perform some operations on a component (e.g., with 
a component in an array of double wells) and arrive at a desired wavefunction. At this point the wavefunction can be copied on the 
other component, and one can gain access (at a later time) 
to this information. In this respect, it would be very 
important to investigate 
the degradation of the quantum switch due to the quantum fluctuations. 

Finally, since the protocol proposed in this paper is not restricted 
to linearly coupled Gross-Pitaevskii equations, one may safely expect 
that - once the effects of quantum corrections are accounted for - it 
can be extended to other sets of coupled equations 
such as the ones arising in the analysis of the weak 
pairing phase of $p$-wave superfluid states of cold atoms \cite{zoller} 
and of quantum wires embedded in $p$-wave superconductors 
\cite{semenoff07}.

{\em Acknoledgements} Several stimulating discussions with E. Fersino, A. Michelangeli, and A. Smerzi are gratefully 
acknowledged. A.T. also thanks L. De Sarlo, 
F. Minardi and C. Fort of the Atomic Physics group at LENS for fruitful 
discussions. In the final stages of the project P.S. benefited from discussions with S. Flach, 
M. Gulacsi, M. Haque and S. Komineas.  P.S. is grateful to the Particle
Theory sector at S.I.S.S.A. for hospitality.  

\appendix

\section{Different potentials and interaction strengths}

When the interaction strengths $g_{ij}$ are different (and, for simplicity, $V_1=V_2$), then 
Eqs. (\ref{lceqa})-(\ref{lceqb}) can be written as
\begin{equation}\label{eqn:stark:gdiff}
i \frac{\partial {\psi}}{\partial t} =
-\frac{1}{2}\Delta {\psi}+ V \psi  + \sum_{j=1}^2 ({\psi}^\dagger G_j \psi) \sigma_j \psi 
+ \alpha(t) P \psi,
\end{equation}
with
\begin{equation}
G_1=\left(
\begin{array}{cc}
g_{11}&0\\
0&g_{12}
\end{array}
\right), ~~
G_2=\left(
\begin{array}{cc}
g_{12}&0\\
0&g_{22}
\end{array}
\right).
\end{equation}
Upon removing the Rabi term in Eq. (\ref{eqn:stark:gdiff}) 
by setting $\psi=U\phi$, with $U$ given by Eq. (\ref{unit}), one gets    
\begin{equation}\label{eqn:stark:gdiff:phi}
i \frac{\partial {\phi}}{\partial t} =
-\frac{1}{2}\Delta {\psi}+ V \phi  + \sum_{j=1}^4 ({\phi}^\dagger \tilde{G}_j \phi) \sigma_j \phi,
\end{equation}
where $\tilde{G_1}={\cal L}_1 - {\cal S}^2 ({\cal L}_1-{\cal L}_2)$ and 
$\tilde{G_2}={\cal L}_2 + {\cal S}^2 ({\cal L}_1-{\cal L}_2)$, where the 
$2 \times 2$ matrices ${\cal L}_{1,2}$ are defined by 
${\cal L}_1=G_1-i {\cal S} \delta_1 {\cal A}$ and 
${\cal L}_2=G_2+i {\cal S} \delta_2 {\cal A}$, with ${\cal A}=-i{\cal S} (\sigma_1-\sigma_2)
+ {\cal C} (\sigma_3 - \sigma_4)$, ${\cal S} = \sin{{\cal I}(t)}$,  ${\cal C} = \cos{{\cal I}(t)}$ and 
$\delta_1=g_{11}-g_{12}$, $\delta_2=g_{22}-g_{12}$. Furthermore 
$\tilde{G}_3=-i{\cal C} {\cal S} ({\cal L}_1-{\cal L}_2)=-\tilde{G}_4$. For 
$g_{11}=g_{12}=g_{22}=g$, then $\delta_1=\delta_2=0$ and $\tilde{G}_1=\tilde{G}_2=G$ and 
$G_3=G_4=0$, so that Eq. (\ref{eqn:nls}) is retrieved.

With different external potentials ($V_1(\vec{r}) \neq V_2(\vec{r})$) and equal interaction 
strengths ($g_{11}=g_{22}=g_{12}=g$), Eqs. (\ref{lceqa})-(\ref{lceqb}) can be written in a matrix form as
\begin{equation}\label{eqn:stark:Vdiff}
i \frac{\partial {\psi}}{\partial t} =
\frac{1}{2} \left( -i \vec{\nabla} \right)^2 {\psi}+\left( {\psi}^\dagger G \psi \right)\psi+V_1({\bf r}) \sigma_1 \psi + V_2({\bf r}) \sigma_2 \psi + 
\alpha(t) P \psi,
\end{equation}
where 
\begin{equation}
\sigma_1=\left(
\begin{array}{cc}
1&0\\
0&0
\end{array}
\right), ~~ 
\sigma_2=\left(
\begin{array}{cc}
0&0\\
0&1
\end{array}
\right), ~~
\sigma_3=\left(
\begin{array}{cc}
0&1\\
0&0
\end{array}
\right) ~~ 
\sigma_4=\left(
\begin{array}{cc}
0&0\\
1&0
\end{array}
\right),
\end{equation}
and $G$ given in Eq. (\ref{psi:G}). It is yet possible to perform a decomposition permitting to 
formally write Eq. (\ref{eqn:stark:Vdiff}) without the Rabi term $\alpha(t) P \psi$. We set  
$\psi=U \phi$, with 
\begin{equation}
U=\left(
\begin{array}{cc}
u_1&u_3\\
u_4&u_2
\end{array}
\right): 
\label{U}
\end{equation}
the Rabi term vanishes, provided that the functions $u_j(\vec{r},t)$ 
($j=1,\cdots,4$) obey the (linear) matrix equation
\begin{equation}
\label{condi}             
i \frac{\partial {U}}{\partial t} = 
-\frac{1}{2}\Delta {U}+\alpha(t) P U + {\cal V}(\vec{r}) 
\left( u_3 \sigma_3 - u_4 \sigma_4 \right), 
\end{equation}
where ${\cal V}(\vec{r})=V_1(\vec{r})-V_2(\vec{r})$.
This matrix equation corresponds to four equations for $u_1$, $u_2$, $u_3$, and $u_4$, which are grouped in two pairs, 
one for $u_1$ and $u_4$, namely
\begin{eqnarray}
i \frac{\partial u _{1}}{\partial t} & = & 
-\frac{1}{2} \Delta u_1 +  \alpha(t) u_{4},
\label{lcequa} \\
i \frac{\partial u_{4}}{\partial t} & = &
-\frac{1}{2} \Delta u_4 - {\cal V}(\vec{r}) u_4 + \alpha(t) u_{1},
\label{lcequb}
\end{eqnarray}
and a similar for $u_2$ and $u_3$ (with ${\cal V}(\vec{r})$ instead of $-{\cal V}(\vec{r})$). 
Eqs. (\ref{lcequa})-(\ref{lcequb}) are two coupled linear Schr\"odinger equations, and the difference between 
the potentials ${\cal V}$ enters as an effective potential in one of the two equations. When ${\cal V}=0$, 
the result $u_1=u_2=\cos{{\cal I}(t)}$ and $u_3=u_4=-i \sin{{\cal I}(t)}$ is readily obtained. 

With the functions $u_j$ defined by Eq. (\ref{condi}), then 
$\phi$ satisfies the equation
\begin{equation}
\label{eqn:stark:Vdiff:phi}
i \frac{\partial {\phi}}{\partial t} =
\frac{1}{2} \left( -i \vec{\nabla}\right)^2{\phi}+\left( {\phi}^\dagger \tilde{G} \phi \right)\phi+V_1({\bf r}) \sigma_1 \phi + V_2({\bf r}) \sigma_2 \phi + 
\vec{{\cal X}} \cdot \vec{\nabla} \phi, 
\end{equation}
with $\tilde{G}=U^\dag G U$ and $\vec{{\cal X}}=\vec{{\cal X}}(\vec{r},t)=-U^{-1} \vec{\nabla}U$. 
Eq. (\ref{eqn:stark:Vdiff:phi}) can be written in the form
\begin{equation}
\label{eqn:stark:Vdiff:phi:2}
i \frac{\partial {\phi}}{\partial t} =
\frac{1}{2} \left( -i \vec{\nabla}+i \vec{{\cal X}} \right)^2{\phi}+\left( 
{\phi}^\dagger \tilde{G} \phi \right)\phi+V_1({\bf r}) \sigma_1 \phi + V_2({\bf r}) \sigma_2 \phi + 
{\cal Y} \phi, 
\end{equation}
where ${\cal Y}=\left( \vec{{\cal X}}^2 - \vec{\nabla} \cdot \vec{{\cal X}} \right)/2$, showing that, with 
different external potentials, 
an effective complex vector potential acts on the multicomponent gas and the effective interaction strengths are both time- and space- dependent 
[since the $\tilde{G}$ depends upon the $U(\vec{r},t)$]. 

\section{Two-Mode Variational Equations}

For the pulse (\ref{timedep}) and at times $t_0 \leq t \leq t_0+\delta$, 
the Lagrangian to be computed is given by 
${\cal L}=\frac{i}{2} \langle \psi_V^\dag \frac{\partial \psi_{V}}{\partial t}- 
\frac{\partial \psi_{V}^\dag}{\partial t} \psi_{V}\rangle-\langle \psi_V^{\dag}\tilde{\cal H} \psi_V\rangle$, 
where 
\begin{equation}
\label{H:tilde}
\tilde{\cal H}=\left(
\begin{array}{cc}
K-\frac{\ell_{11}}{2} \vert \psi_{v1} \vert^2 -\frac{\ell_{12}}{2} \vert \psi_{v2} \vert^2 & \gamma\\
\gamma& K-\frac{\ell_{12}}{2} \vert \psi_{v1} \vert^2 -\frac{\ell_{22}}{2} \vert \psi_{v2} \vert^2
\end{array}
\right), 
\end{equation}
$\psi_V$ is defined in Eq. (\ref{variational}) and $K=-\frac{1}{2} \frac{\partial^2}{\partial x^2}$. Then, one gets
\begin{displaymath}
{\cal L}=-N_1 \dot{\varphi}_1-N_2 \dot{\varphi}_2-\gamma \sqrt{\frac{\ell_{22}}{\ell_{11}}}
F\left( \frac{\ell_{22}}{\ell_{11}} \right) \sqrt{N_1 N_2} \cos{(\varphi_1-\varphi_2)}
-\frac{\ell_{11}^2}{24} N_1-\frac{\ell_{22}^2}{24} N_2+ \mu+
\end{displaymath}
\begin{equation}
\label{Lag}
+\frac{\ell_{11}^2}{12} N_1^2 +\frac{\ell_{22}^2}{12} N_2^2 + \frac{\ell_{22}}{8 } \ell_{12} 
L\left( \frac{\ell_{22}}{\ell_{11}} \right) N_1 N_2,
\end{equation}
where the functions $F(\theta)$ and $L(\theta)$ are defined as
\begin{equation}
\label{F:L}
F(\theta) \equiv \int_{-\infty}^{\infty} \frac{dy}{\cosh{(y)} \cosh{(\theta y)}}; \, \, \, 
L(\theta) \equiv \int_{-\infty}^{\infty} \frac{dy}{\cosh^2{(y)} \cosh^2{(\theta y)}}.
\end{equation}
Note that $F(0)=\pi$, $F(1)=2$, $L(0)=2$, and $L(1)=4/3$, while both functions $\to 0$ as $\theta \to \infty$. 
For small values of $\delta \theta=\theta-1$, 
one has $F(\theta)\approx 2 -\delta\theta+\frac{24-\pi^2}{36} \delta\theta^2$ 
and $L(\theta)\approx \frac{4}{3} -\frac{2}{3} \delta\theta-\frac{2(\pi^2-15)}{45} \delta\theta^2$. For 
$\ell_{11}=\ell_{22}$, from Eq. (\ref{Lag}) one gets, apart from constant terms, the Lagrangian
\begin{equation}
{\cal L}=-N_1 \dot{\varphi}_1-N_2 \dot{\varphi}_2-2 \gamma \sqrt{N_1 N_2} \cos{(\varphi_1-\varphi_2)}
+\frac{\ell_{11}^2}{12} N_1^2+\frac{\ell_{22}^2}{12} N_2^2+\frac{\ell_{11} \ell_{12}}{6} N_1 N_2.
\label{Lag:s}
\end{equation}

Introducing the variables (\ref{variables}), the equations of motions for $\eta$ and $\varphi$ obtained from the Lagrangian 
(\ref{Lag}) are
\begin{equation}
\left \{ \begin{array}{ll}
\dot{\eta}=2 \gamma' \sqrt{1-\eta^2} \sin{\varphi}, \\
\dot{\varphi}=-2 \gamma' \frac{\eta}{\sqrt{1-\eta^2}}\cos{\varphi}+\Delta E + M \eta,
        \end{array}
\right.
\label{var:t:compl}
\end{equation}
where
\begin{equation}
\gamma'=\frac{\gamma}{2} \sqrt{\frac{\ell_{22}}{\ell_{11}}}
F\left( \frac{\ell_{22}}{\ell_{11}} \right) ,
\end{equation}
\begin{equation}
M=\frac{1}{12} 
\left[ \frac{3}{2} \ell_{12} \ell_{22} L \left( \frac{\ell_{22}}{\ell_{11}} \right) -\ell_{11}^2 -\ell_{22}^2\right], 
\label{mass}
\end{equation}
and 
\begin{equation}
\Delta E=\frac{1}{24} \left( \ell_{22}^2 - \ell_{11}^2\right).
\label{Delta_E}
\end{equation}
The variables $\eta, \varphi$ are canonically conjugate 
dynamical ones with respect to the effective Hamiltonian
\begin{equation}
\label{Hamiltonian-pend}
H_{eff} = \frac{M}{2} \eta^2 + 2 \gamma' \sqrt{1-\eta^2} \cos{\varphi}+ \Delta E \cdot \eta. 
\end{equation} 
Equation (\ref{Hamiltonian-pend}) is the Hamiltonian of a non-rigid pendulum, whose mass and length depend 
on the parameters $\ell_{ij}$. For $\ell_{11}=\ell_{22}$, then $M=\ell_{11}(\ell_{12}-\ell_{11})/6$, 
$\Delta E=0$ and $\gamma'=\gamma$, retrieving Eq. 
(\ref{var:t}) and the effective Hamiltonian (\ref{Hamiltonian-pend-sempl}). 
When $\ell_{12}=\ell_{11}=\ell_{22}$, then the mass of the pendulum is vanishing. When $\ell_{11} \neq \ell_{22}$, 
then the presence of the detuning $\Delta E$ favours self-trapping and can be studied as in \cite{raghavan99}.



\begin{thebibliography}{99}
\bibitem{nota1} Notice that one can pass from $\alpha<0$ to $\alpha>0$ simply by shifting the phase difference 
$\chi=\chi_2-\chi_1$ between the condensate wavefunctions 
$\psi_j$ (where $\psi_j=\sqrt{N_j}e^{i \chi_j}$) of $\pi$: $\chi \to \chi+\pi$.
\bibitem{footnote} Using the notation $U=e^b$, with the time-dependent matrix $b(t)$ defined by 
$b=-i \int_0^t \tilde{P}(t') dt'$, we observe that when the matrices $b$ and 
$\partial b/\partial t$ commute, then the decomposition 
$\psi=U \phi$ allows to cast 
Eq. (\ref{eqn:stark:3}) in the homogeneous form.
\end{thebibliography}
\end{document}